\documentclass[
    reprint,
    preprintnumbers,
    superscriptaddress,
    nofootinbib,
     amsmath,amssymb,
     aps,
     prd,
    floatfix,
    ]{revtex4-2}

    \usepackage{graphicx}
    \usepackage[usenames,dvipsnames]{xcolor} % colour
    \usepackage{times, mathptmx} % Times font
    \usepackage[utf8]{inputenc}
    \usepackage{color}
    \usepackage{hyperref}
    \usepackage[normalem]{ulem} 
    \usepackage{physics}
    \usepackage{comment}
\usepackage{booktabs, multirow}
    \usepackage{enumitem}
    \usepackage{comment}
    \usepackage{bm}
    \usepackage{bigints}

\hypersetup{
  colorlinks   = true, %Colours links instead of ugly boxes
  urlcolor     = blue, %Colour for external hyperlinks
  linkcolor    = blue, %Colour of internal links
  citecolor   = blue %Colour of citations
}
    \allowdisplaybreaks[1]
    \graphicspath{{figures/}}

\usepackage{float}

\newcommand{\qmul}{Geometry, Analysis and Gravitation, School of Mathematical Sciences, Queen Mary University of London,
Mile End Road, London E1 4NS, United Kingdom}
\newcommand{\potsdam}{Institut f{\"u}r Physik und Astronomie, Universit{\"a}t Potsdam, Haus 28, Karl-Liebknecht-Str. 24/25, 14476, Potsdam, Germany}
\newcommand{\aei}{Max Planck Institute for Gravitational Physics (Albert Einstein Institute), Am M{\"u}hlenberg 1, Potsdam 14476, Germany}

\begin{document}

\title{Numerical simulations of Scalar Dark Matter Around Binary Neutron Star mergers}

\author{Rohan Srikanth}
\email{rohan.srikanth@uni-potsdam.de}
\affiliation{\potsdam}
\author{Tim Dietrich}
\affiliation{\potsdam}
\affiliation{\aei}
\email{tim.dietrich@uni-potsdam.de}
\author{Katy Clough}
\email{k.clough@qmul.ac.uk}
\affiliation{\qmul}
%\affiliation{\oxford}

\begin{abstract}
Binary neutron star mergers provide a laboratory for probing fundamental physics through their gravitational-wave emission and electromagnetic counterparts. In particular, they may allow us to explore signatures of physics beyond the Standard Model in strong-gravity regimes, such as those of dark matter. In this work, we investigate the dynamics of light dark matter, modeled as a minimally coupled scalar field, surrounding a binary neutron star system. Our primary focus is to assess whether the scalar field remains bound to the binary over the late inspiral-merger timescales and to determine its potential impact on observable signatures. We find that, in a range of scenarios, the scalar field forms a common cloud around the binary that does not disperse. At sufficiently high densities, this leads to measurable effects, including a dephasing of the binary inspiral, a less compact post-merger remnant, and suppression of the dynamical ejecta. For densities motivated by astrophysical considerations, however, these effects remain small and are unlikely to be detectable with current or next-generation gravitational-wave observatories. 
\end{abstract}

\maketitle

\section{Introduction}
The detection of the binary neutron star (BNS) merger GW170817 \cite{LIGOScientific:2017vwq}, together with its electromagnetic counterparts—the short gamma-ray burst GRB170817A, its afterglow \cite{LIGOScientific:2017zic}, and the kilonova AT2017gfo \cite{LIGOScientific:2017pwl, Lipunov:2017dwd, Shappee:2017zly, J-GEM:2017tyx, Goldstein:2017mmi}—has revolutionized our understanding of neutron star mergers by providing direct constraints on the dense-matter equation of state (EOS).
A key objective in the study of BNS systems is to determine their fundamental properties—masses, spins, and tidal deformabilities-and to use these to probe the EOS governing matter at supranuclear densities.
Gravitational-wave (GW) observations, X-ray observations with NICER \cite{Miller:2019cac, Miller:2021qha, Riley:2019yda, Riley:2021pdl, Raaijmakers:2019dks} and radio timing of massive pulsars such as PSR~J0348+0432 with a mass of $2.01 \pm 0.04 \ M_{\odot}$ \cite{Antoniadis:2013pzd} and PSR~J0740+6620 ($2.08 \pm 0.07\ M_{\odot}$) \cite{Fonseca:2021wxt} provide complementary probes, offering additional constraints on the EOS of neutron stars.

The above analyses assume that the BNS is in a vacuum, but in principle, it is also possible that the system is surrounded by dense environments such as accretion discs, gas, or dark matter. In this work, we examine the case of dark matter (DM) environments around BNS mergers. We study the extent to which it accumulates around the binary during the inspiral phase, and its effect on the GW waveform, matter ejecta, and post-merger effects. There have been several works highlighting the potential to use compact objects as DM probes, see \cite{Bertone:2019irm, Bramante:2023djs, deLavallaz:2010wp, Baryakhtar:2022hbu, Bertone:2007ae} and references therein. Future GW observations will have much greater precision and could be sensitive to dephasing in the waveforms induced by environmental/matter effects in a binary black hole (BBH) or BNS merger event \cite{Bertone:2019irm}. 

The effects of DM can be explored by broadly classifying it into two categories according to its mass. For particle-like DM, in the mass range $m_{\rm DM}\sim {\rm eV}-{\rm GeV}$, numerous studies have examined its effect inside neutron stars (NSs) and its overall impact on BNS mergers \cite{Kumar:2025ytm, Routaray:2024lni, Liu:2025qco, Luo:2025psd, Giangrandi:2022wht, Ivanytskyi:2019wxd, Sagun:2021oml, Kouvaris:2010vv, Kouvaris:2007ay, DelPopolo:2020hel, Nelson:2018xtr, Bauswein:2020kor}; see \cite{Grippa:2024ach} for a recent review. In \cite{Grippa:2024ach, Kouvaris:2007ay}, they outline the rate of accumulation of DM particles within NSs, which depends on the asymptotic energy density, scattering cross-section with the Standard Model (SM) particles, and the age of the NSs. For a single neutron star, without self-interactions, it is of the order of $10^{-10} \ M_{\odot}$ over a time scale of $t = 10 \ \mathrm{Gyr}$ \cite{Ellis:2017jgp}, but this estimate may increase for heavy weakly interacting massive particles (WIMPs) and is enhanced by Bondi accretion near Galactic centers up to 0.01\% of the star’s total mass \cite{Ivanytskyi:2019wxd}. While the accumulation is enhanced for BNS compared to individual NSs, the DM fraction depends on the ambient DM energy density. At higher DM densities, this fraction is relatively suppressed for BNSs because dynamical friction accelerates the inspiral, causing the system to merge earlier, such that it spends less time in the DM environment \cite{Takatsy:2025nnw}. Once accreted within the NSs, DM particles can transfer kinetic energy to the SM particles during their infall at semi-relativistic speeds, consequently overheating the neutron star \cite{Bramante:2023djs, Baryakhtar:2020gao, Kouvaris:2007ay}. Further constraints are provided by considering that WIMPs, if accreted substantially, can form a black hole due to their own gravitational collapse \cite{Kouvaris:2010vv, Kouvaris:2007ay}. 

If the DM mass is $\lesssim 30{\rm eV}$, standard DM densities imply large occupation numbers, meaning it is bosonic in nature, and may be effectively described by a classical field. For the spin-0 case such DM can be modeled using the Klein-Gordon equation \cite{Hui:2016ltb, Hu:2000ke, Schive:2014dra, Niemeyer:2019aqm, Grin:2019mub, Marsh:2015xka, Brito:2022lmd}, for a review, see \cite{Ferreira:2020fam, Urena-Lopez:2019kud}. DM in this mass range may also be spin-1 (vectors) \cite{Farzan:2012hh, Hancock:2025ois}, or pseudoscalars \cite{Kim:1986ax}, but the scalar case is the simplest, and most well motivated by high energy extensions to the SM. Low-mass DM models have historically been motivated as a possible solution to the cusp-core problem \cite{Bullock:2017xww, deBlok:2009sp, Robles:2012uy} and the missing satellite problem \cite{Perivolaropoulos:2021jda}, providing a modification to the usual $\Lambda\text{CDM}$ model in the small-scale regime \cite{Ferreira:2020fam, Perivolaropoulos:2021jda}, although these differences in cosmological structure between simulations and observations may be explained by unmodelled baryonic effects \cite{DelPopolo:2021bom, Del_Popolo_2009}. 
Probes of this mass range include laboratory searches such as the CASPEr experiments—probing axion-like particles (ALPs)—as well as cosmological constraints from CMB spectral distortions \cite{Marsh:2015xka}.
Light dark matter is often termed “wave dark matter", with the de Broglie wavelength setting its characteristic length scale \cite{Hui:2021tkt}. The effect of wave DM influences the compact objects only if the characteristic\footnote{Strictly speaking, one should consider the de Broglie wavelength, but since in our simulations the velocities are relativistic and spatially varying, we usually refer to the Compton wavelength, which is a constant and of a similar order.} wavelength is of the order of or smaller than the distance between the binaries.
If the wavelength of wave DM is much larger, the gradient pressure of the field resists accumulation, preventing the formation of overdensities and smoothing the DM profile. Wave DM interacts gravitationally with the binary, and due to its coherent nature and pressure support on the scale of its wavelength, it can accumulate around black holes (BHs). In contrast, as shown in \cite{Merritt:2003qk, Bertone:2005hw, Merritt:2002jz}, particle-like DM tends to scatter and disperse due to self-annihilations and kinetic heating. In particular, in the context of equal-mass binary black hole (BBH) mergers, N-body simulations have demonstrated that particle DM overdensities tend to disperse \cite{Bertone:2005hw, Kavanagh:2018ggo}. For BNS mergers, however, particle-like DM modeled as a non-interacting fermionic fluid, coupled gravitationally with the neutron star matter, can survive the merger and continue to influence post-merger dynamics \cite{Giangrandi:2025rko}. The behavior of an extended environment of wave DM as studied for BHs in \cite{Zhang:2022rex, Choudhary:2020pxy, Bamber:2022pbs, Aurrekoetxea:2023jwk,Tomaselli:2024ojz,Guo:2025pea,Roy:2025qaa} is largely unexplored for BNS. This motivates a study to investigate the fate and the effect of wave DM in the post-merger phase. 
In this work, the DM mass is set by the characteristic length scale of the problem—of the order of the binary separation—leading us to focus on the $10^{-11} - 10^{-12} \ \rm eV$ range. 

Numerical-relativity simulations have been employed in various studies to explore the impact of DM on BBH mergers as well as BNS mergers and post-merger dynamics. The effects of wave DM on BBH mergers have been studied both numerically \cite{Zhang:2022rex, Choudhary:2020pxy, Bamber:2022pbs, Aurrekoetxea:2023jwk, Guo:2025pea, Cheng:2025wac} and semi-analytically \cite{Tomaselli:2024ojz, Guo:2025pea, Roy:2025qaa}. In \cite{Aurrekoetxea:2023jwk}, they investigated the impact of wave DM on the GW waveform of BBH mergers. Their results indicated dephasing of order $\mathcal{O}(10\%)$, depending on the initial binary separation and assuming initial asymptotic energy densities of order $\rho_\varphi = 10^{-7} - 10^{-9}\ \rm M^{-2}$. Other related works have included self-interactions \cite{Aurrekoetxea:2024cqd}, studied the effects of non-trivial kinetic terms \cite{Machet:2025vzt}, and quantified the dynamical friction \cite{Traykova:2023qyv,Traykova:2021dua}, and Magnus effects \cite{Wang:2024cej}, aiming to understand the evolution and dynamics of the scalar field profile in various astrophysical environments.
A central question involves the fate of the DM environment during the merger and post-merger, and its interaction with the binary. Ref.~\cite{Giangrandi:2025rko} investigated the impact of particle-like DM, modeled as a non-interacting fermionic fluid, on BNS mergers during both inspiral and post-merger phases. They showed that the presence of DM modifies the compactness of the post-merger remnant depending on the binary mass. Considering DM mass fractions of $0.5\%$ and $3\%$ relative to the baryonic component--defined as
$f = \frac{M_{\mathrm{DM}}}{M_{\mathrm{tot}}}$ with $M_{\mathrm{tot}}$ the total gravitational mass of the binary--they found that for higher-mass systems ($\sim 2.8\,M_\odot$) DM can favor a prompt collapse to a black hole. In these cases, the inclusion of DM shortens the lifetime of the post-merger remnant.
Several studies have considered how the presence of particle-like DM within the NSs can modify their internal structure and their EOS \cite{Ivanytskyi:2019wxd, Giangrandi:2022wht, Abac:2021txj, Panotopoulos:2017idn, Hippert:2022snq, Diedrichs:2023trk}. Additionally, self-interacting DM has been studied in \cite{Guver:2012ba, Karkevandi:2021ygv, Giangrandi:2022wht}. In \cite{Giangrandi:2022wht}, it was shown that self-interactions can soften the EOS of the NSs, and the fraction of bosonic self-interacting DM was constrained to be $f < 5 \%$ based on the mass-radius relation and the tidal deformability. Another study considered the mirror DM model, conducting single star (TOV) tests and BNS simulations, suggesting that the presence of DM reduces the lifetime of the remnant for $f = 5\% , \  10 \%$ \cite{Emma:2022xjs}. 

Early studies of DM examined its impact on GW signals from BNS mergers \cite{Ellis:2017jgp}. Using an approximate mechanical model of the BNS system, they demonstrated that DM embedded within NSs can alter the waveform, potentially generating an additional peak in the GW spectral density. The DM fraction considered in their work $f \sim 5 - 10 \%$ is much higher than realistic values (see discussion below), but might be achieved through additional couplings, either via the dark sector or beyond SM physics \cite{Foot:2004pa, Fan:2013yva}. Further, \cite{Bezares:2019jcb} employed numerical simulations to study BNS mergers with a bosonic DM model, with the baryonic matter modeled as a perfect fluid and DM as a complex scalar field. Their results indicated the emergence of a strong $m=1$ mode in the GW waveform. However, no significant impact on the overall dynamical evolution or the GW waveform signal was observed. Here as well, the fraction of the bosonic DM considered was higher than expected with $f \sim 5 \%, \ 10 \%$. The effect on tidal deformability of a neutron star admixed with scalar field DM was studied in \cite{Diedrichs:2023trk}, which also examined how self-interactions of the field modify both the tidal deformability and the EOS. Several studies have also explored the effects of additional matter couplings with nucleonic matter \cite{Hook:2017psm, Kumamoto:2024wjd}.

In the absence of self-interactions or additional couplings with baryonic matter, asymptotic wave DM energy densities are likely to be very small and the impact on merger dynamics will be correspondingly small. Using the results for BHs \cite{Hui:2019aqm} as a proxy, one can approximate the density profile of the scalar DM near a single neutron star as
\begin{equation}
    \rho(r) \sim  10^{-31} 
\left( \frac{M_{\rm NS}}{M_\odot} \right)^2 
\left( \frac{\rho_i}{1 \, {\rm GeV/cm}^3} \right) 
\left( \frac{r_i/r_s}{10^6} \right)^{3/2},
\end{equation}
where $\rho_i$ denotes the ambient DM density, $r_s$ the Schwarzschild radius associated with the star, and $r_i$ is the radius of influence\footnote{The profile will differ inside the NSs and at the boundary, but we have checked the profile for a single neutron star, and $\sim r^{3/2}$ scaling provides a good approximation.}. While additional mechanisms, such as superradiance, can enhance DM overdensities around spinning BHs \cite{Brito:2015oca, Baryakhtar:2020gao, Hui:2022sri, Siemonsen:2022ivj, Cardoso:2004nk, East:2017ovw}, their effect on NSs is generally smaller due to the weaker coupling between the field and the spinning object in most of the simplest axion models \cite{Cardoso:2015zqa, Day:2019bbh, Chadha-Day:2022inf}, unless additional couplings are invoked \cite{Cardoso:2017kgn}. 

In this study, we focus on two key questions: (i) \textit{How does a wave DM environment influence the merger and post-merger dynamics of BNSs?} and (ii) \textit{How does the DM grow and accumulate around the binary, and is it retained to the merger?}
As discussed above, our work considers a much denser asymptotic scalar environment than expected in realistic astrophysical settings, and so even the small dephasing we detect is a best-case scenario. The results for the accumulation of the DM should, however, rescale straightforwardly to lower, realistic densities, allowing us to understand whether the overdensity persists, and by how much compared to its initial value. 

This paper is organized as follows. First, we describe the scalar field setup and initial profiles, along with the initial data procedure and equations involved in dynamical evolution in Sec~\ref{sec:setup}. In Sec~\ref{sec:results}, we discuss our results observed in the GW waveform, ejecta properties, and post-merger effects. We then summarise and conclude in Sec~\ref{sec:discussion}. The convergence test results and details of the multigrid setup are in Appendix~\ref{sec:Appendix} and \ref{subsecMG}. Throughout this paper, we use geometric units with $M_{\odot} = G = c = 1$.

\begin{figure*}[t]
    \centering
    \includegraphics[width=\textwidth]{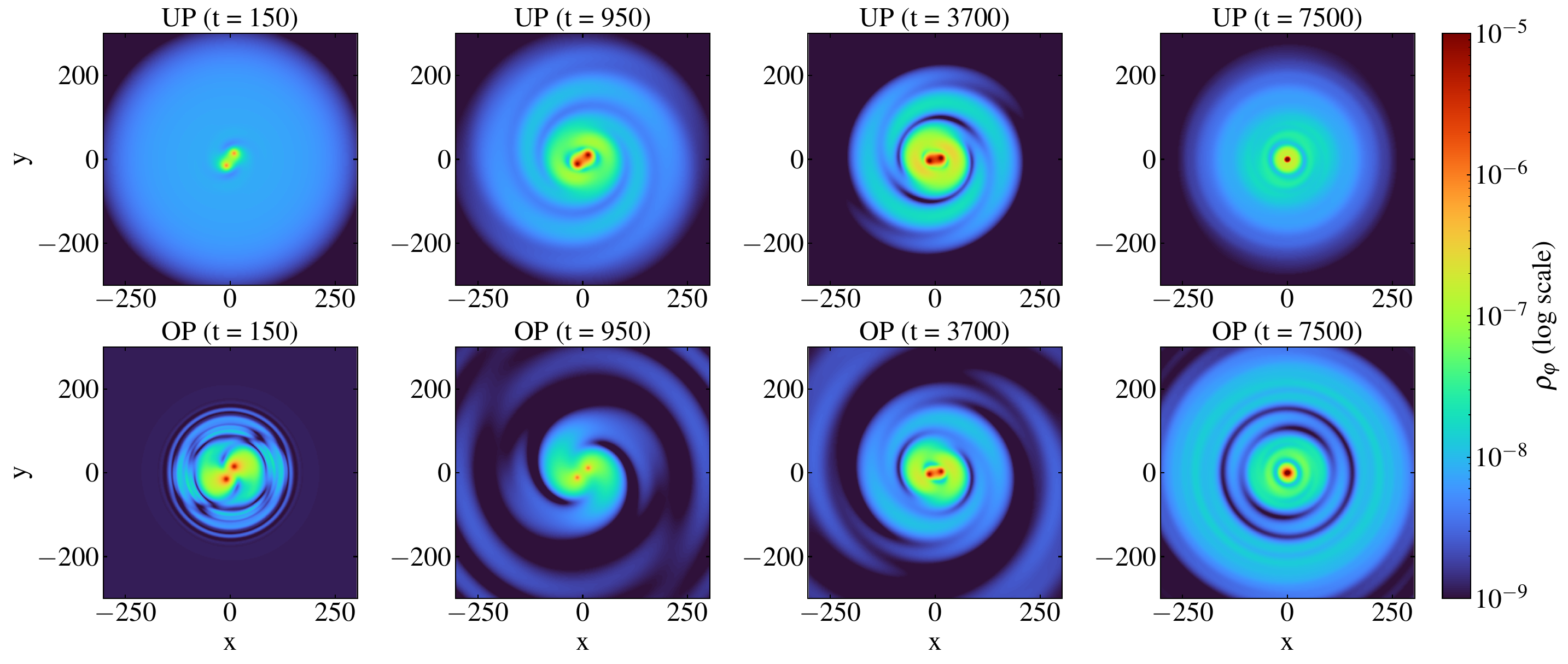}
    \caption{2D schematics showing the comparison of the energy densities for the two scalar field initial profiles. Top row: For the \textit{Uniform profile} (UP) at different times starting from early-inspiral till the post-merger phase. Bottom row: Energy densities snapshots for the \textit{Overdense profile} (OP) at the same time stamps of the simulation, showing localized enhancement close to the binary even at $t=150$.  The UP is effectively a nearly homogeneous field initially, as the cut-off is set at a large radius. At later times, a co-rotating configuration develops, giving rise to characteristic spiral-arm features in the energy density.} 
    \label{fig:profiles}
\end{figure*}
\section{Framework and Setup}\label{sec:setup}
In this section, we discuss the scalar field model and initial profiles, followed by describing the initial data for the BNS and dynamical evolution using BAM. 
\subsection{Scalar field model and profiles}
As discussed above, the scalar field can be treated in the classical regime 
for the $10^{-11} - 10^{-12} \ \rm eV$ mass range we study
\cite{Hui:2021tkt}. 
We consider a minimally coupled complex scalar field $\varphi$ along with the baryonic matter contribution for the NSs described by the action,
\begin{align}\label{eqaction}
    S &= \int d^4x \ \sqrt{-g} \left( \mathcal{L}_\varphi + \mathcal{L}_{NS} \right) \\ 
    \nonumber  & = \int d^4x \ \sqrt{-g} \left(\frac{R}{16\pi} - \frac{1}{2} (\nabla_{\mu} \varphi) (\nabla^\mu \varphi^*) - \frac{1}{2}\mu^2 \varphi \varphi^*  + \mathcal{L}_{NS} \right),
\end{align}
where $\varphi^*$ denotes the complex conjugate of the field, {$\mathcal{L}_{NS}$ is the Lagrangian of the baryonic matter, the corresponding sources and the conservation equations are discussed in \ref{subsubsec:NS}}. $\mu$ is the inverse length scale related to the Compton wavelength as $\mu = \frac{2\pi}{\lambda_c} = \frac{m_s c}{\hbar}$, where $m_s$ is the scalar field mass. Here, we consider a simple quadratic potential without self-interactions; the dynamics of the scalar field are governed by the massive Klein-Gordon equation. For details regarding the evolution, we refer to the subsection~\ref{subsec:KG}.

In the following, we explore two classes of initial conditions for the scalar field profile around the BNS to study the dependence of the evolution on the chosen initial configuration. Although the quantitative results are indifferent, neither profile initially represents a quasi-stationary configuration. We find that both induce long-lived oscillations in the scalar field that depend strongly on the chosen initial data, motivating further work to improve the initial data, similar to the ones studied in \cite{Siemonsen:2023age, Nee:2024bur}.

\subsubsection{Uniform profile}
The first class of initial profiles is motivated by \cite{Bamber:2022pbs}, where the evolution of the scalar field was studied in the context of BBHs. This profile corresponds to a nearly constant density configuration with a smooth cutoff, ensuring that the field approaches an asymptotic value at spatial infinity. For sufficiently large cutoffs, it effectively mimics a homogeneous background field. The explicit form is given by
\begin{equation} \varphi_r(r) = \frac{1}{2}\varphi_0 \left[ 1 + \tanh\left(\frac{r_0 - r_c}{b} \right) \right], 
\end{equation}
where \( r_c = \sqrt{x^2 + y^2 + z^2} \) is the radial coordinate, \( b \) controls the width of the transition. $\varphi_0$ is the central uniform field amplitude, and we set it to be \( 5 \times 10^{-4} \), and $\varphi_r$ refers to the real part of the complex scalar field decomposed as $\varphi = \varphi_r + i\varphi_i$. This setup effectively acts as a reservoir from which the scalar field can accrete onto the binary over a few orbits. To facilitate this, the cut-off radius \( r_0 \) and the width parameter \( b \) are chosen appropriately, and for a large enough radius, this profile resembles a homogeneous $\varphi_r(r) \sim \varphi_0$ profile. Once the DM is accreted close to the binary due to gravitational interaction, it forms a quasi-equilibrium co-rotating configuration as seen in the energy density snapshots in Fig~\ref{fig:profiles}.

\subsubsection{Overdense profile}
Over the preceding evolutionary timescales, the scalar field may have developed an overdensity around the stars.
Motivated by this, we approximate such a configuration by the profile
\begin{equation}
    \varphi_r(r) = \left[ A_1 \exp\left( -\frac{(r - \mu_1)^2}{2\sigma_0^2} \right) + b \right] \cdot \exp\left(-\alpha (r - r_d)\right) - c,
\end{equation}
where \( \mu_1 \), \( \sigma_0 \) control the position and width of the central Gaussian, \( r_d \) sets the onset of the exponential tail, \( \alpha \) defines the decay rate, and \( A_1 \), \( b \), \( c \) control the amplitude and asymptotic offset.
The parameters are chosen to match the radial field profile through one star when evolving the Uniform profile for a few orbits and fitting the real part of the scalar field. We tested multiple parameter values and observed qualitatively similar behavior in the evolution. In Fig.~\ref{fig:profiles}, we present snapshots of the scalar field energy density $\rho_\varphi$ at different simulation times. At initial times, the \textit{Uniform profile} appears nearly homogeneous within the extraction region due to its large cut-off radius, whereas the \textit{Overdense profile} exhibits localized enhancement. After dynamical evolution, the scalar field settles into a corotating
configuration accompanied by the emission of scalar radiation in the form of outgoing waves.
\subsection{Initial data and Numerical implementation}
The initial configurations for NSs are constructed using \texttt{SGRID} \cite{Tichy:2009yr, Tichy:2012rp, Dietrich:2015pxa, Tichy:2019ouu}, a pseudo-spectral code
that uses surface fitting coordinates to solve the Einstein constraint equation through the extended conformal thin sandwich (XCTS) formalism \cite{York:1998hy, Pfeiffer:2002iy}. We employ an eccentricity–reduction procedure following \cite{Tichy:2019ouu}, in order to mitigate residual orbital eccentricity after the construction of the initial data. This approach is motivated by the findings of \cite{Bamber:2022pbs}, wherein they observed that overlaying a scalar field on top of a binary system (as is the case for the \textit{Uniform profile}) can induce additional eccentricity, in particular through the introduction of a spurious linear momentum. After applying the eccentricity–reduction procedure, we achieve a residual eccentricity of order $\sim 4\times10^{-3}$. Following this, the dynamical evolution is performed with BAM \cite{Bruegmann:2006ulg, Thierfelder:2011yi, Dietrich:2015iva, Bernuzzi:2016pie, Dietrich:2018bvi, Schianchi:2023uky, Neuweiler:2024jae, Gieg:2024jxs,Neuweiler:2025klw}. BAM evolves the metric in time using the methods-of-line approach, utilizing finite-difference stencils for spatial discretization. For the study, we use the 4th order Runge-Kutta integration with a Courant-Friedrichs-Lewy (CFL) value of 0.25. Further, we also utilize the Z4c constraint damping terms \cite{Bona:2003fj, Bona:2005pp, Gundlach:2005eh, Bernuzzi:2009ex, Hilditch:2012fp}, and a $1+\log$ slicing condition \cite{Alcubierre:2002kk}. BAM's infrastructure consists of a hierarchy of refinement levels with adaptive mesh refinement capabilities (AMR), and we apply the Berger-Oliger scheme \cite{Berger:1984zza} for the local-time stepping. For the inner refinement levels, the boxes can move and are automatically adjusted during the evolution, which helps in narrowing down and tracking the stars. For our simulations, in the nested Cartesian boxes, we employ a total of 7 refinement levels with 3 inner levels that move along with the stars. Reflecting boundary conditions are imposed along the \( z \)-axis, while Sommerfeld (radiative) boundary conditions are used on the remaining boundaries. The matter evolution consists of two parts: the general relativistic hydrodynamical (GRHD) evolution for the NSs and the scalar field evolution. 
\subsubsection{Source terms: NSs}{\label{subsubsec:NS}}
For the case of NSs, we use the  perfect fluid approximation, resulting in the stress-energy tensor,
\begin{equation}
    T^{\text{NS}}_{\mu\nu} = \rho h \ u_{\mu} u_{\nu} + p \ g_{\mu\nu},
\end{equation}
where $\rho$ is the density, $p$ the pressure, $h$ is the enthalpy, and $u_{\mu}$ is the 4-velocity. We can then determine the evolution of the matter variables with the help of conservation equations, taking the form,
\begin{subequations}
    \begin{align}
    \nabla_{\mu} T^{\mu\nu} = 0, \label{eq:EMcons} \\
    \nabla_{\mu} (\rho u^{\mu}) = 0, \label{eq:Bcons}  \\
    P(\rho, \epsilon) = p, \label{eq:EOScons} 
\end{align}
\end{subequations}
where Eq.~\eqref{eq:EMcons} is the local energy-momentum conservation, Eq.~\eqref{eq:Bcons} is the baryon number conservation, and Eq.~\eqref{eq:EOScons} is the EOS of the baryonic matter pertaining to the NSs. For our study, the baryonic matter is modeled using a piecewise-polytropic representation of the SLy EOS \cite{Douchin:2001sv}. For the NSs, the evolution includes the GRHD equations. We begin with a set of primitive variables $\textbf{w} = (\rho, v_i, \epsilon)$, which are the rest-mass density, the fluid velocity, and the internal energy density. Here, the quantities are measured according to the Lagrangian observer, and we further introduce a set of conservative variables $\textbf{q} = (D, S_i, \tau)$ --  the conserved rest mass, momentum, and internal energy density of the Eulerian observer. The primitive and conservative variables are related by
\begin{subequations}\label{eq:cons-vars}
\begin{align}
D   &= \rho W,\\
S_i &= \rho h W^2 v_i,\\
\tau&= \rho h W^2 - p - D,
\end{align}
\end{subequations}
with \( W = (1 - v^2)^{-1/2} \). Rewriting Eq.~(\ref{eq:EMcons}), Eq.~(\ref{eq:Bcons}), Eq.~(\ref{eq:EOScons}), as a first-order, flux-conservative, hyperbolic system, we get
\begin{equation}\label{eq:flux-cons}
\frac{1}{\sqrt{-g}}
\left(
  \partial_{0}\!\left(\sqrt{\gamma}\,\mathbf{q}\right)
  + \partial_{i}\!\left(\sqrt{-g}\,\mathbf{F}^{\,i}\right)
\right)
= \mathbf{S},
\end{equation}
where
\begin{subequations}\label{eq:defs}
\begin{align}
\mathbf{q}(\mathbf{w}) &= (D,\, S_j,\, \tau), \\[2pt]
\mathbf{F}^{\,i}(\mathbf{w}) &= \Big(
 D\!\left( v^i - \tfrac{\beta^i}{\alpha} \right),\;
 S_j\!\left( v^i - \tfrac{\beta^i}{\alpha} \right) + p\,\delta^{\,i}{}_{j},\;
 \tau\!\left( v^i - \tfrac{\beta^i}{\alpha} \right) + p\,v^i
\Big), \\[2pt]
\mathbf{S}(\mathbf{w}) &= \Big(
0,\;
T^{\mu\nu}\!\left(\partial_\mu g_{\nu j} - \Gamma^{\sigma}_{\nu\mu} g_{\sigma j}\right), \notag \\[2pt]
&\qquad\;
\alpha\!\left( T^{\mu 0}\,\partial_\mu \log\alpha - T^{\mu\nu}\Gamma^{0}_{\nu\mu} \right)
\Big).
\end{align}
\end{subequations}

\subsubsection{Scalar field}\label{subsec:KG}
In the case of the scalar field, the energy-momentum tensor can be derived from the action Eq.~(\ref{eqaction}), and is given by,
\begin{align}\label{eqemfield}
T_{\mu\nu} = \tfrac{1}{2} \Big[ & \, \nabla_\mu \varphi \nabla_\nu \varphi^* 
+ \nabla_\nu \varphi \nabla_\mu \varphi^*  \nonumber \\
& - \tfrac{1}{2} \Big( g^{\mu\nu} \nabla_\mu \varphi \nabla_\nu \varphi^* 
+ \tfrac{1}{2} \mu^2 \varphi \varphi^* \Big) \Big].
\end{align}
Klein-Gordon equations are second-order in nature, but we can rewrite the equations in first-order in time form using the time-derivative of the field
\begin{subequations}
\begin{align}
\Pi &\equiv \frac{1}{\alpha} \left(\dot{\varphi} - \beta^i \partial_i \varphi \right),
\end{align}
with the evolution equations for the field and its time derivative being,
\begin{align}
\partial_t \varphi &= \alpha \Pi + \beta^i \partial_i \varphi, \\[1ex]
\partial_t \Pi &= \beta^i \partial_i \Pi
+ \alpha \left(
    \gamma^{ij} \partial_i \partial_j \varphi
   - \gamma^{ij} \Gamma^{k}{}_{ij}\, \partial_k \varphi
   + K \Pi
   - \frac{dV(\varphi, \varphi^*)}{d\varphi_0^2} \varphi
  \right) \nonumber \\
&\quad + \gamma^{ij} (\partial_j \alpha)\, \partial_i \varphi .
\end{align}
\end{subequations}
where $\varphi_0^2 = |\varphi|^2$, and $K$ is the trace of the extrinsic curvature $
K_{ij} = -\frac{1}{2\alpha} \left( \partial_t \gamma_{ij} - D_i \beta_j - D_j \beta_i \right)$. The time derivative of the field $\Pi$ is initialised such that $\Pi_r(t=0) = 0$ and $\Pi_i(t=0) = -i\mu \varphi_r(t=0)$.

\subsubsection{Multigrid}
After superposing the scalar field on the BNS, we re-solve the Hamiltonian and momentum constraints using the in-built multigrid solver in BAM~\cite{Galaviz:2010mx,Moldenhauer:2014yaa,Dietrich:2018bvi}. This ensures consistency with Einstein’s equations by minimizing constraint violations, which is essential since the spacetime metric is altered by the addition of the scalar field. It is important to note that we do not re-solve the equations for GRHD or the scalar field equations (Klein-Gordon), which are necessary to obtain fully consistent, quasi-stationary initial configurations.
The multigrid algorithm uses a recursive method of iteratively solving discretized equations with constant interaction between a hierarchy of grids \cite{Brown:2004ma, Moldenhauer:2014yaa}. On each grid, the discretized equations are solved by relaxation; the fine grids capture the shorter wavelength effects, while the coarser one captures the long-wavelength effects. These are communicated using the restriction and prolongation operators. Further details are provided in Appendix~\ref{subsecMG}.

\subsubsection{Diagnostics}
\begin{figure}
    \centering
    \includegraphics[width=\linewidth]{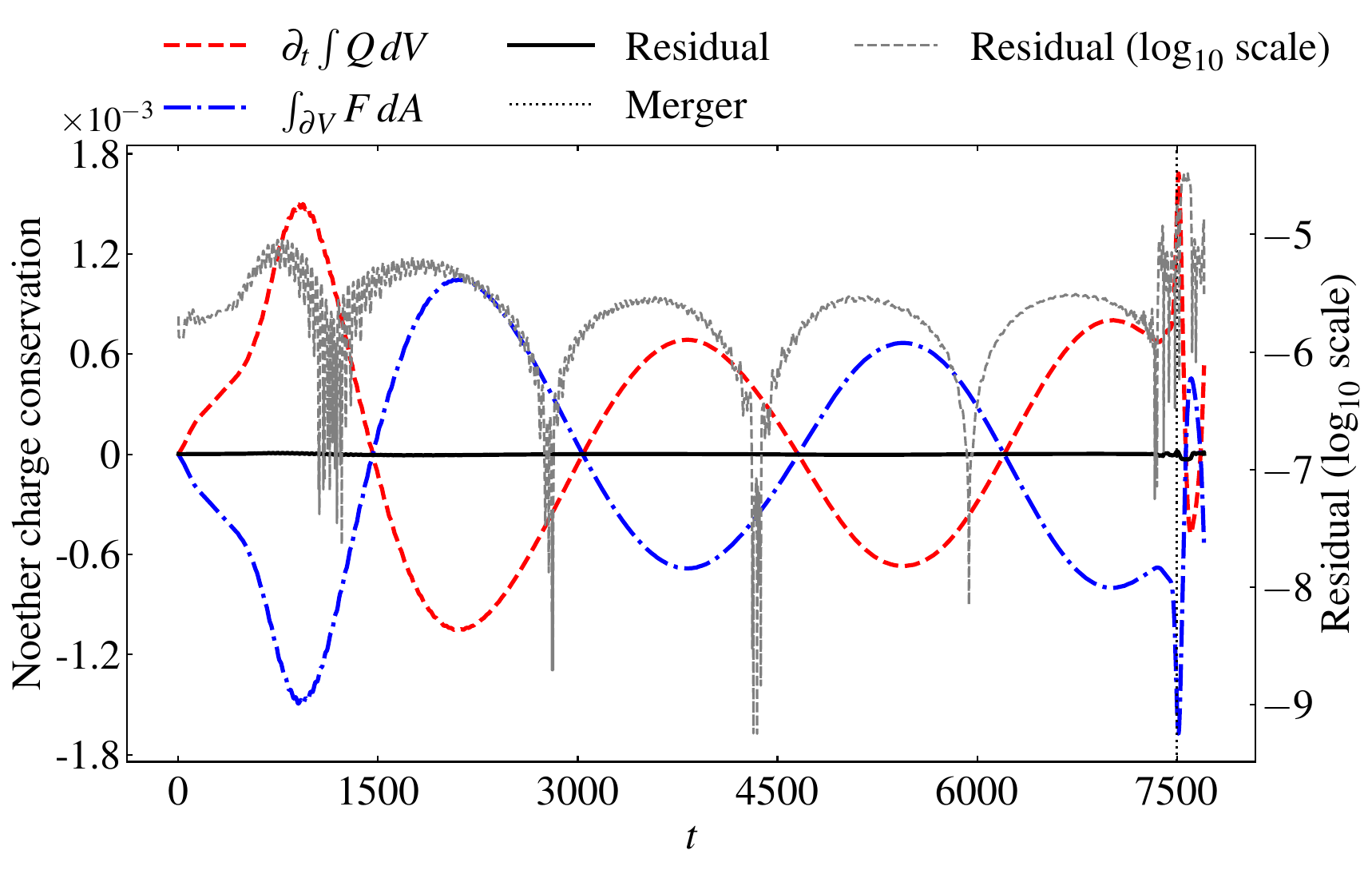}
    \caption{Noether charge conservation plot for the $2.7$ total mass binary for the \textit{Uniform profile} with the scalar field mass $\mu = 0.085$, extracted at a radius of $\rm R=40$. The red dotted lines represent the time derivative of the charge ($\int Q \ dV$), the blue dashed line is the flux ($\int F \ dA $), both are oscillatory, and the residual is shown in a black solid line and is conserved throughout the simulation. The axis on the right shows the residual on a log scale.}
    \label{fig:noether}
\end{figure}

As diagnostics, we monitor two primary quantities: (i) the Noether charge of the scalar field, and (ii) Angular momentum exchange between the binary and the scalar, following \cite{Clough:2021qlv, Croft:2022gks}. These diagnostics serve both to validate the correctness of the simulations and to provide insight into the dynamics of the field–binary interaction. \newline 
For the Noether charge, we check if the rate of change of a charge in an enclosed volume is equal to the flux across the boundary of the volume, such that,
\begin{equation}
    \partial_t \int_{\Sigma} Q_{\phi} \ dV = \int_{d\Sigma} F\cdot\hat{n} \ dA \label{eqdiag},
\end{equation}
where the flux is given by $F = \sqrt{h} \ J^{\mu} N_{\mu}$ \cite{Croft:2022gks}, $N_\mu = (N_0, N_i) = (N_i \beta^i, N_i)$ is the coordinate normal vector, $h$ is the determinant on the extracted surface. $N_i$ in Cartesian coordinates is given in terms of the unit coordinate vector $s_i = \frac{x_i}{r}$ as $N_i = \frac{\gamma_{ij} s^j}{\sqrt{\gamma_{ij} s^i s^j}}$.
Checking the validity of Eq.~\eqref{eqdiag} ensures that the simulation is sufficiently resolved and the results are consistent. In Fig.~\ref{fig:noether}, the conservation of the Noether charge is demonstrated for the $2.7$ total mass binary for the \textit{Uniform profile}. The plot shows that the rate of change of the charge (red line) and the flux (blue line) are in balance. Consequently, the residual, defined as the difference between the left- and right-hand sides of Eq.~\eqref{eqdiag} and shown by the black solid line (for linear scale) and in gray dotted line (for a log scale), is conserved throughout the simulation.
\newline
For the angular momentum conservation, assuming that we have an isolated system in an asymptotically flat region, one can split the ADM contributions of the system into matter and curvature, and quantify the exchange between the binary and the scalar field. We tracked the angular momentum terms for both the scalar field and the NSs, to check for conservation by computing the respective charges and fluxes, following \cite{Clough:2021qlv}. Details of the formalism and diagnostic plots, along with their convergence, are in Appendix~\ref{sec:Appendix}.

\begin{table}[t]
\centering
\small 
\setlength{\tabcolsep}{2.2pt} 
\renewcommand{\arraystretch}{1.45}  
\caption{Table showing the simulation parameters. From left to right: The total mass of the binary, the initial scalar field profile, the identifier for each configuration that is being used throughout the paper, parameter related to the mass of the scalar field $\mu$ in $M_{\odot}$ units, the initial separation $d_{\text{BNS}}$  of the binary in $M_{\odot}$ units, and finally the resolution at the moving AMR levels. The asymptotic energy density of the field for both profiles is set to be $\rho_\varphi \sim 1 \times 10^{-8}$. We also, for reference, ran simulations without the scalar field (NF) for both binary masses.}
\vspace{0.3cm}  
\begin{tabular*}{\linewidth}{@{\extracolsep{\fill}} l l l c c l}
\toprule
\addlinespace[0.1cm] 
\textbf{Total Mass} & \textbf{Profile} &\textbf{Identifier} & \textbf{$\mu$} & \textbf{Init.\ Sep.} & \textbf{Resolution} \\
 & & & & \textbf{$d_{{BNS}}$} & \\
\addlinespace[0.1cm]  
\midrule
\addlinespace[0.1cm] 
\multirow{4}{*}{\textbf{2.7}}
 & Uniform & $\rm UP^{2.7}_{0.17}$   & 0.17  & 18 & 128, 144, 192 \\[0.1cm]
 & Uniform & $\rm UP^{2.7}_{0.085}$  & 0.085 & 18 & 144 \\[0.1cm]
 & Overdense &$\rm OP^{2.7}_{0.17}$ & 0.17  & 18 & 128, 144, 192 \\[0.1cm]
 & Vacuum & NF                             & ---   & 18 & 128, 144, 192 \\
\addlinespace[0.15cm]  
\midrule
\addlinespace[0.1cm] 
\multirow{3}{*}{\textbf{3.6}}
 & Uniform & $\rm UP^{3.6}_{0.17}$   & 0.17 & 21 & 144 \\[0.1cm]
 & Overdense & $\rm OP^{3.6}_{0.17}$ & 0.17 & 21 & 144 \\[0.1cm]
 & Vacuum & NF                              & ---  & 21 & 144 \\
\addlinespace[0.1cm] 
\bottomrule
\end{tabular*}
\label{tab:simparams}
\end{table}
\begin{figure}
    \centering
    \includegraphics[width=\linewidth]{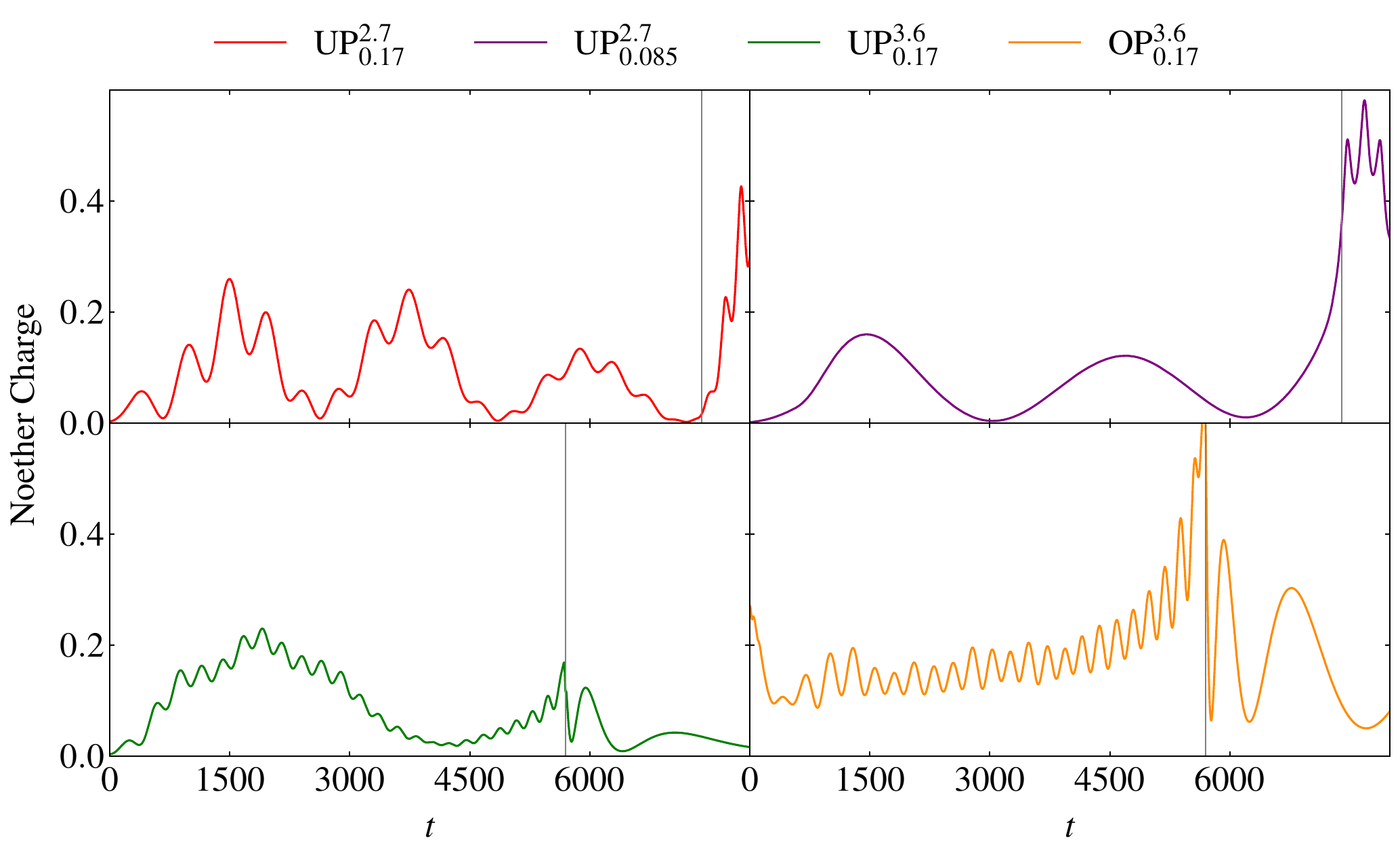}
    \caption{Noether charge distribution around a single star part of the binary for different scalar field profiles extracted at a radius of $\rm R=10$. Top left: $M_{\rm tot} = 2.7$, for the \textit{Uniform profile} and $\mu = 0.17$ ($\rm UP^{2.7}_{0.17}$). Top right: For the \textit{Uniform profile} and a lower field mass $\mu = 0.085$ ($\rm UP^{2.7}_{0.085}$). Bottom left: $M_{\rm tot} = 3.6$ and \textit{Uniform profile} ($\rm UP^{3.6}_{0.17}$). Bottom right: $M_{\rm tot} = 3.6$ and \textit{Overdense profile} ($\rm OP^{3.6}_{0.17}$). The black solid line indicates the merger time. We note that the scalar field oscillations form a \textit{beating} pattern for the $\mu = 0.17$, and are smoother for the lower mass case on the top right owing to the gradient pressure. For the higher mass binary, the oscillation frequency is higher for both profiles.}
    \label{fig:allnoether}
\end{figure}
\begin{figure*}[!t]
    \centering
    \includegraphics[width=0.7\linewidth]{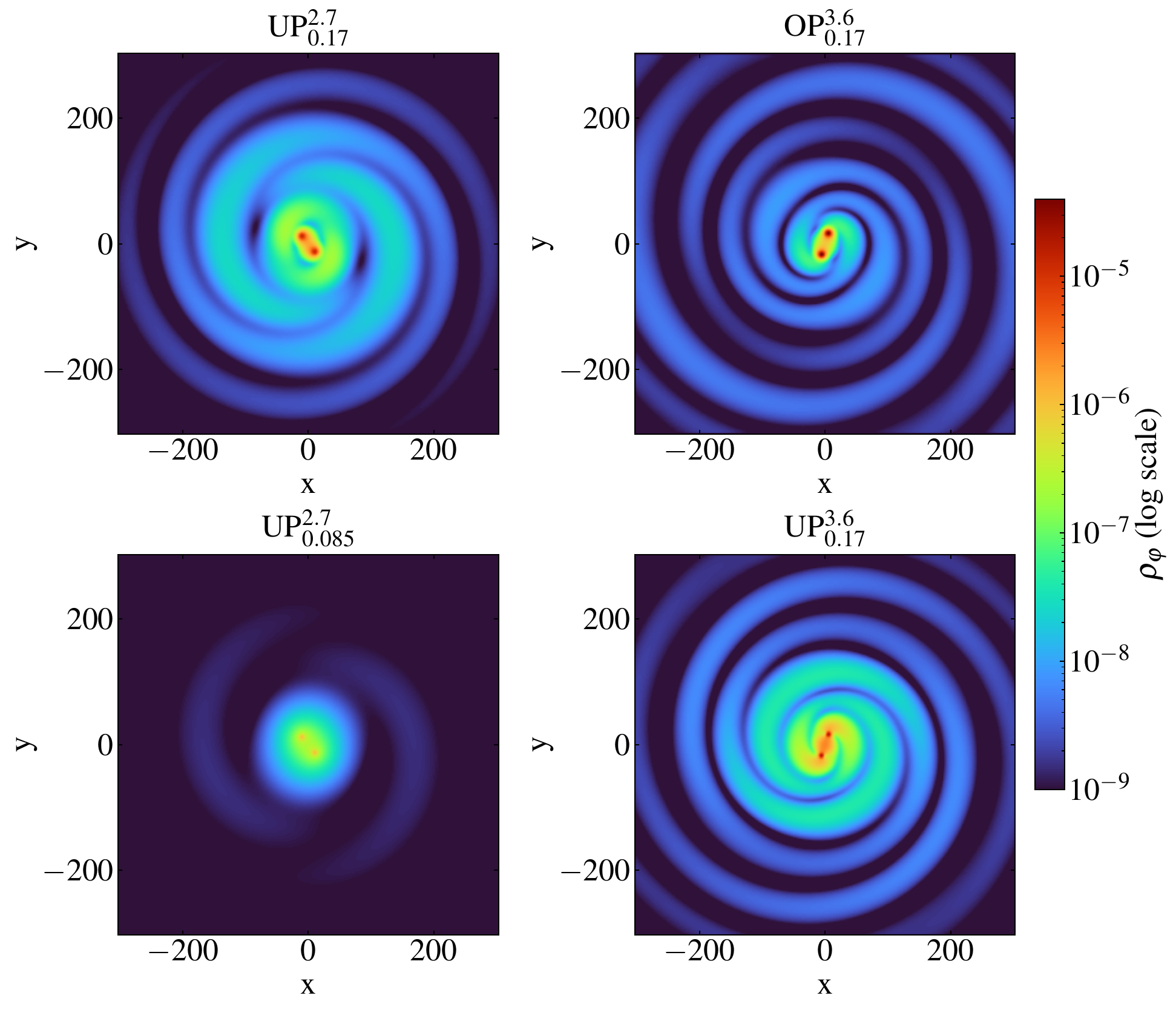}
    \caption{2D snapshots at $t=2000$ of the scalar–field energy density $\rho_{\varphi}$. 
Top-left: \textit{Uniform profile}, $M_{\rm tot}=2.7$, $\mu=0.17$. 
Top-right: \textit{Overdense profile}, $M_{\rm tot}=3.6$, $\mu=0.17$. 
Bottom-left: \textit{Uniform profile}, $M_{\rm tot}=2.7$, $\mu=0.085$. 
Bottom-right: \textit{Uniform }profile, $M_{\rm tot}=3.6$, $\mu=0.17$. 
The lower-mass field ($\mu=0.085$) shows markedly reduced central accumulation, consistent with a stronger gradient pressure. The $3.6$ mass binary simulations exhibit broader, more extended spiral-arm features, potentially due to higher oscillations of the scalar field and a deeper gravitational potential well.}
    \label{fig:profileslowandhigh}
\end{figure*}
\section{Results}\label{sec:results}
Below, we outline the results beginning with the evolution of the scalar field profiles, the GW waveform, followed by the effect of the scalar field on matter ejecta and on the fate of the remnant.
\subsection{Evolution of the DM profiles}
Our study includes a suite of seven simulations to investigate the effects of different binary total masses and scalar field profiles on BNS mergers. We explored binaries with two different total masses $M_{\rm tot} = 2.7, 3.6$. For each, we used two scalar field profiles: \textit{Uniform profile} and \textit{Overdense profile}, both with a field mass of $\mu = 0.17$. To explore the impact of the scalar field mass, we conducted an additional simulation with a lower field mass of $\mu = 0.085$ for the \textit{Uniform profile} and $M_{\rm tot} = 2.7$. We choose the scalar mass \(\mu = 0.17\) so that the
(Compton) wavelength is comparable to the initial binary separation \(d_{BNS}\) (or to an integer multiple of it, \(\lambda_c \simeq n\,d_{BNS}\) \cite{Tomaselli:2024ojz}). A massive scalar mediates a
Yukawa interaction \(\propto e^{-r/\lambda_C}/r\) with range
\(\lambda_c\). Thus, if \(\lambda_c \ll d_{BNS}\) (\(\mu d_{BNS} \gg 1\)), we move closer to particle-like DM regime, while,  if \(\lambda_C \gg d_{BNS}\) (\(\mu d_{BNS} \ll 1\)), the field is effectively long-ranged on the orbital scale and the binary is subjected to only weak gradients, reducing the efficiency of angular-momentum exchange.
Choosing \(\lambda_C \sim d_{BNS}\) (\(\mu d_{BNS} \sim \mathcal{O}(1)\)) maximizes the interaction between the binary and the field. Hence, exploring different scalar field masses allows us to investigate the influence of the scalar field's pressure term on the gravitational collapse and dynamics of the system. For comparison, we also ran two simulations without the scalar field. In our setup for both profiles, the asymptotic field amplitude is initially $\phi \sim 10^{-4}$, resulting in an initial energy density of approximately $\rho_\phi \sim 10^{-8}$. Details of the specific simulation parameters are provided in Table~\ref {tab:simparams}. 

To study the field dynamics, we examine the evolution of the Noether charge around one of the stars in the binary as it orbits, shown in Fig.~\ref{fig:allnoether}.\footnote{Because of the symmetry of the binary system, the results are quantitatively identical if we consider the other star.} In the top-left panel, we show the evolution for the \textit{Uniform profile} with scalar field mass $\mu = 0.17$ for $M_{\rm tot} = 2.7$. The evolution displays oscillations with a characteristic \textit{beating} pattern that persists until merger.
\begin{figure*}[!t]
    \centering
    \includegraphics[width=\linewidth]{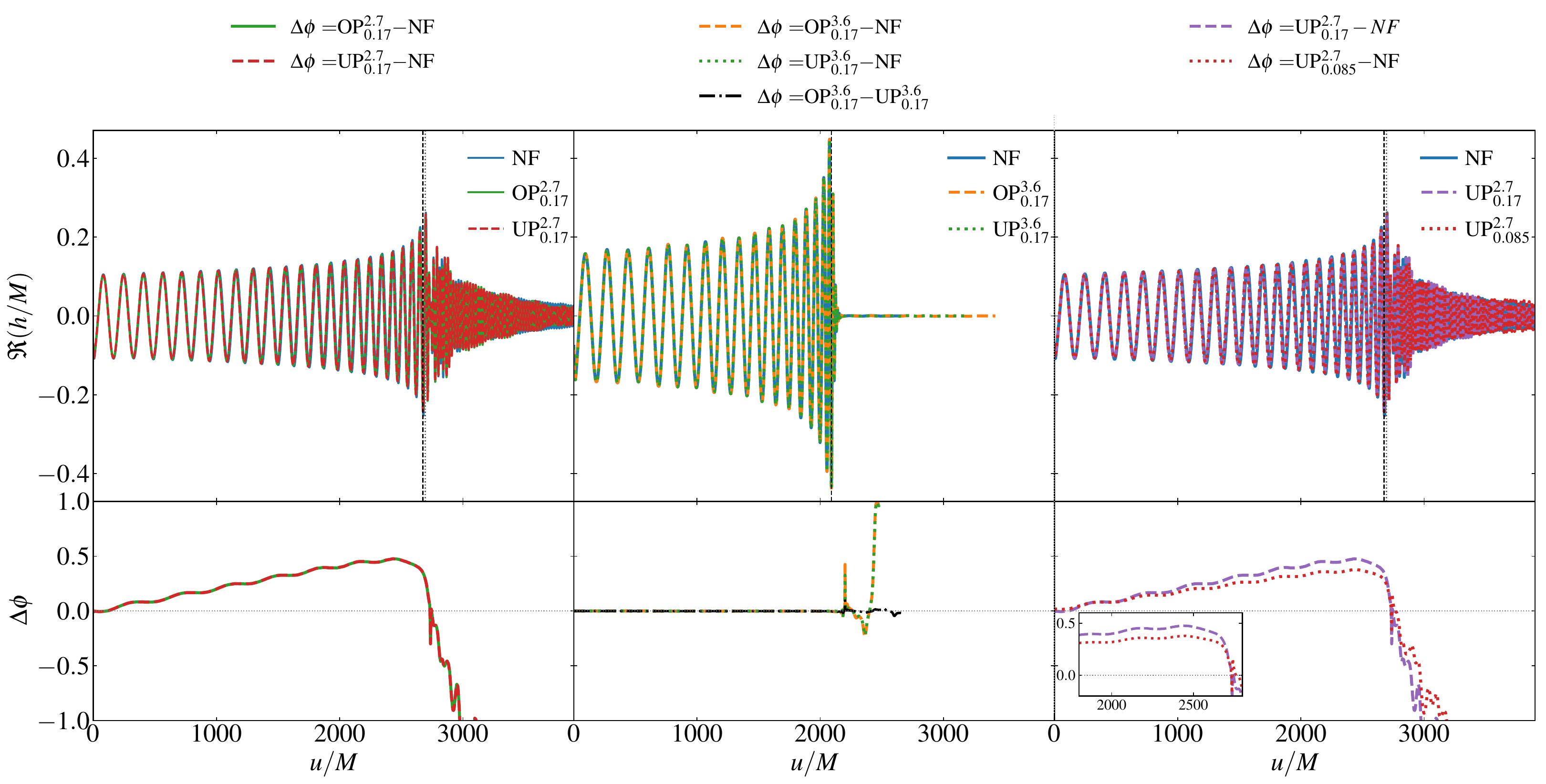} 
    \caption{Left: The real part of the GW strain (top panel) and the corresponding phase differences (bottom panel) for different scalar field profiles extracted at a radius $\rm R_{ext} =  1000$. The \textit{Uniform} (UP) and \textit{Overdense profiles} (OP) both show the same dephasing with the vacuum case (NF) of roughly $0.3$ rad close to the merger. Middle: The real part of the GW strain (top panel) and the corresponding phase differences (bottom panel) for different scalar field profiles with a total mass of $M_{\rm tot} = 3.6$, along with the NF of the same $M_{\rm tot}$. Only a negligible dephasing is observed between the vacuum case and the configurations with the scalar field. Right: The real part of the GW strain (top panel) and the corresponding phase differences (bottom panel) for two different masses, $\mu = 0.17$ and $\mu = 0.085$. The lower scalar field mass exhibits a smaller accumulated dephasing relative to the vacuum case close to the merger, owing to the pressure term, as shown in the inset.}
    \label{fig:gw_lowmass}
\end{figure*}
\newline
After the merger, the field amplitude increases sharply as the system forms a dense hypermassive neutron star. In the top-right panel, we present the case of a lower scalar field mass. Here, the evolution is comparatively smooth since the gradient pressure dominates and suppresses overdensities. Figure~\ref{fig:noether} shows the Noether charge and flux oscillating, implying that the scalar field flows in and out of the central volume in waves, characteristic of a long-wavelength oscillating mode being excited. In this regime, the post-merger phase shows substantial amplification.
Lowering the scalar mass further may reduce this enhancement. The bottom-left and bottom-right panels correspond to the $M_{\rm tot} = 3.6$ binary, evolved with the \textit{Uniform} and \textit{Overdense profiles}, respectively. In this case, the oscillation frequency is higher, and it is interesting to note that even after the prompt collapse of the binary, the scalar field persists. The \textit{Uniform profile} exhibits accumulation and decay patterns similar to the lower-mass binary, whereas the \textit{Overdense profile} produces oscillations of nearly constant amplitude that gradually increase near merger. This is likely due to a specific mode excitation at $t=0$, arising from the construction of the initial data for which the system does not begin in the ground state of the field, as explained above. 

In Fig.~\ref{fig:profileslowandhigh}, we present 2D snapshots of the scalar field energy density profiles in logarithmic scale for different binary masses and scalar field masses. This is extracted after the system has settled into a configuration wherein the field and the binary co-rotate. The red colored regions indicate the position of the binary, also corresponding to the overdense region with an enhancement of about $10^4$ times that of the initial asymptotic value for both profiles, and this is further amplified post-merger, as the scalar field energy density reaches $\rho_{\varphi} \sim 10^{-3}$. Hence, although the Noether charge exhibits oscillations as discussed above, the scalar field remains mostly bound throughout the simulation timescale, and the local density is enhanced after the merger. There is also an accumulation between the binary owing to DM being trapped in the gravitational potential well. The top right and bottom right panels correspond to the higher-mass binary, where the scalar field develops a much broader distribution with pronounced "spiral-arm" features. This behavior is potentially linked to the higher oscillation frequency of the field and the deeper gravitational potential of the binary. The top left panel shows the \textit{Uniform profile} for $\mu = 0.17$, while the bottom left panel corresponds to $\mu = 0.085$, both for $M_{\rm tot} = 2.7$. In the lower scalar mass case, the overdensity is substantially reduced due to the enhanced gradient pressure, which inhibits scalar field accumulation and favors dispersion compared to the $0.17$ case, with an enhancement compared to the asymptotic value of the order $10^2 - 10^3$. 

\begin{figure*}[!t]
    \centering
    \includegraphics[width=\linewidth]{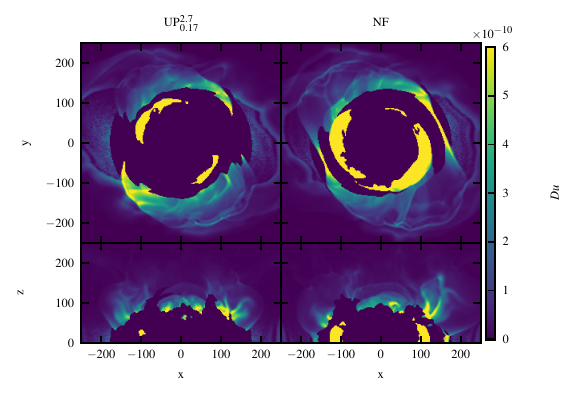}
    \caption{2D snapshots of unbound ejecta for the vacuum (NF) case and the \textit{Uniform profile} ($\rm UP^{2.7}_{0.17}$) with scalar field mass $\mu = 0.17$ for $M_{\rm tot} = 2.7$ binary extracted at an AMR refinement level $l=3$. Top panel: Comparison in the equatorial plane, which is typically dominated by shock-driven ejecta, showing a clear suppression in the presence of the scalar field. 
Bottom panel: Corresponding snapshots along the polar plane, confirming the overall reduction of the ejecta due to the scalar field. All snapshots are taken at $t = 8800$.  
}
    \label{fig:2dhydroa}
\end{figure*}
\begin{figure}
    \centering
    \includegraphics[width=\linewidth]{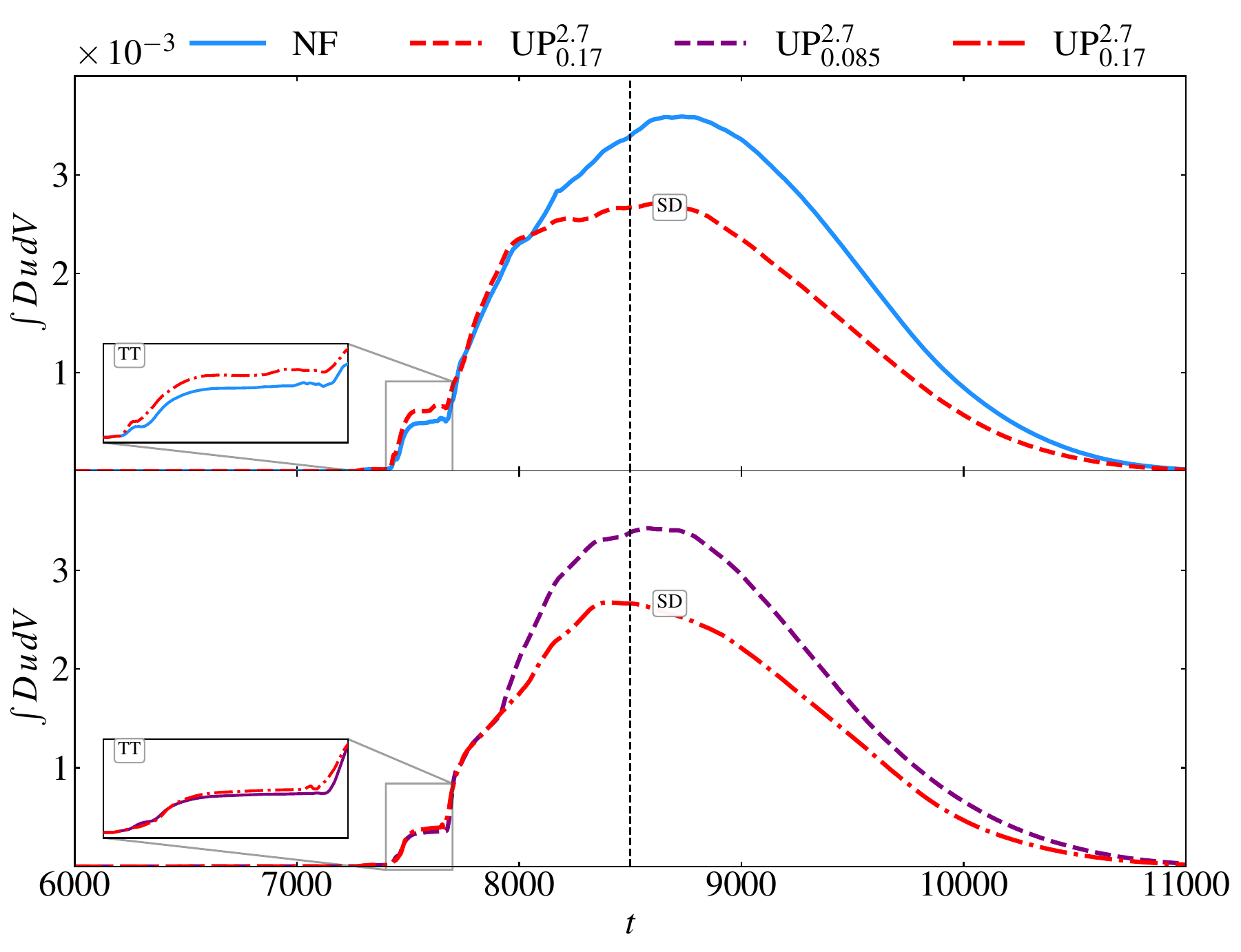}
    \caption{Volume-integrated unbound ejecta mass at the AMR level $l=3$. Top panel: Comparison between the $\rm UP^{2.7}_{0.17}$ and the vacuum (NF) case. Regions dominated by tidal and shock-driven ejecta are marked, with insets zooming in on the tidal-tail contribution. The scalar field enhances the tidal-tail ejecta while suppressing the shock-driven component, resulting in a net reduction of total ejecta.  Bottom panel: Comparison of the same for two scalar field masses, $\rm UP^{2.7}_{0.17}$ and $\rm UP^{2.7}_{0.085}$. Reducing the scalar field mass shifts the behavior closer to the vacuum case, increasing the shock-driven ejecta while reducing the tidal-tail contribution.}
    \label{fig:0dhydroa}
\end{figure}
\begin{figure}
    \centering
    \includegraphics[width=\linewidth]{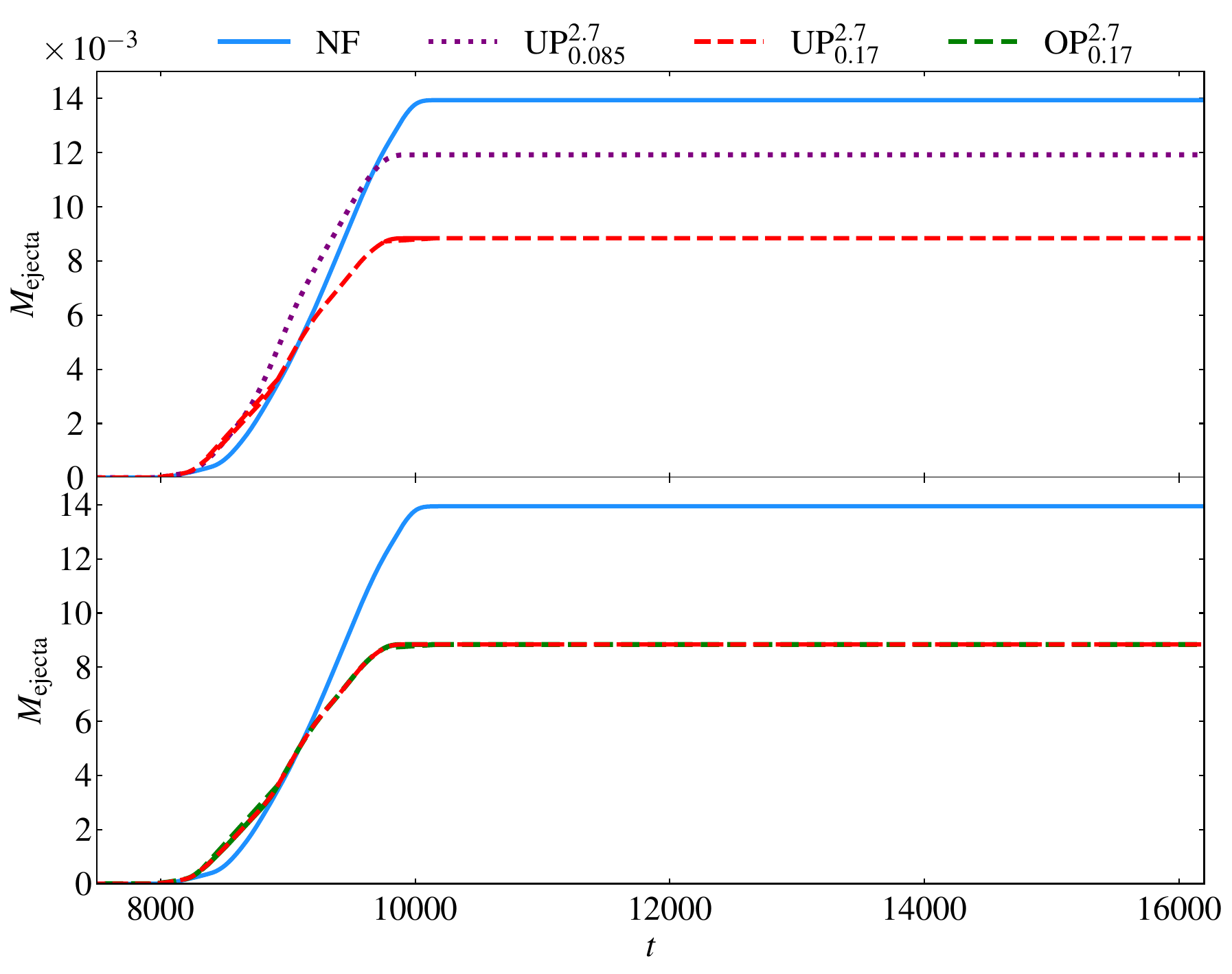}
    \caption{Unbounded ejecta mass comparison for different scalar field configurations at extraction radius $\rm R = 300$ and at the finest level. Top panel: Comparison between two scalar field masses, $\mu = 0.085$  ($\rm UP^{2.7}_{0.085}$) and $\mu = 0.17$ ($\rm UP^{2.7}_{0.17}$). Reducing the scalar field mass decreases the accumulation of energy density, driving the dynamics closer to the vacuum (no-field) case. Bottom panel: Comparison for a fixed scalar field mass, $\mu = 0.17$, across two different initial profiles ($\rm UP^{2.7}_{0.17}$) and ($\rm OP^{2.7}_{0.17}$) and the vacuum case (NF), we notice roughly $60 \%$ reduction in presence of the scalar field, relative to the vacuum. }
    \label{fig:ejecta}
\end{figure}

\subsection{Gravitational Waves: Dephasing}

We extract the GW signals from the simulations at different extraction radii $r_{\rm{ext}}$, using the retarded time in terms of the retarded time coordinate given by 
\begin{equation}
     u = t - R_{ext} - 2M_{ADM} \ln \left(\frac{r_{ext}}{2M_{ADM}}-1\right),
\end{equation}
with the extraction radius $\rm R_{ext}$.
We extract the dominant quadrupolar strain mode, $h_{2|2|}$, for all simulations and present the results in Fig.~\ref{fig:gw_lowmass} at an extraction radius of $\rm R_{ext} = 1000$. 
The left panels compare two initial scalar field profiles--- \textit{Uniform} (UP) and \textit{Overdense profile} (OP)---together with the vacuum case (NF) for the same $M_{\rm tot} = 2.7$. In the top panel is the waveform strain, and the lower panel, the corresponding phase difference, $\Delta\phi(t) \equiv \phi_{\rm DM}(t)-\phi_{\rm vac}(t)$.
Two main features to note: (i) the UP and OP profiles yield indistinguishable phasing within our numerical accuracy throughout the inspiral, and (ii) both produce a net dephasing of $\sim 0.3~\rm{rad}$ at merger relative to the vacuum simulation (NF).  
The dephasing is possibly due to angular-momentum extraction by the scalar field through scalar radiation, in agreement with previous analyses \cite{Aurrekoetxea:2023jwk}. 
As in the BBH case, we quantify the angular-momentum transfer and associated source terms (see Appendix~\ref{sec:Appendix}), but with one caveat in the case of BNSs. The NSs also contribute to the angular momentum exchange due to the nonzero stress-energy tensor; this complicates the one-to-one mapping between the cause of dephasing and the angular momentum source term. Consequently, our interpretation is qualitative—albeit consistent with the diagnostics. As illustrated in the Appendix~\ref{sec:Appendix}, in the middle panel of Fig.~\ref{fig:allham}, increasing the numerical resolution systematically suppresses the angular momentum contributions associated with the stars. Given that our simulations cover only the last few orbits before merger, and since the dephasing at an earlier stage of the inspiral can not be addressed by our simulations, we can not definitively quantify if the observed dephasing is large enough to be seen with GW detectors, however, the rather large field values and relative small dephasing might indicate that observing such a dephasing is unlikely with current and next generation experiments \cite{Abac:2025saz}, in addition, it is possibly degenerate with other matter effects.

In the middle panel, we show the $(2,2)$-mode strain and the corresponding phase difference for the $M_{\rm tot} = 3.6$. 
Across the inspiral, we find \emph{negligible dephasing} with respect to the NF case (here with $M_{\rm tot}=3.6$) that we attribute to the higher stellar compactness and the correspondingly shorter inspiral: both effects limit the build–up of scalar overdensities and reduce the net torque exerted by the field on the binary. 
The merger leads to prompt collapse, so no long–lived post–merger effects are present, and the waveform shows no additional post–merger features.  Despite this, the simulation with a scalar field does merge \emph{slightly earlier} than its vacuum counterpart; however, the accumulated dephasing is significantly smaller than in the $M_{\rm tot}=2.7$ case.
\newline
In the right panel, we compare two scalar field masses. 
The lower mass case, $\mu = 0.085$, exhibits a \emph{smaller dephasing} with respect to the NF run than the higher scalar field mass case.  
The reduction in $|\Delta\phi|$ for $\mu=0.085$, is expected as it is consistent with the weaker scalar overdensity, due to the gradient pressure resisting accumulation, and in turn, suppresses angular-momentum transfer between the binary and the field. This is shown in the inset, which is zoomed in close to the merger for clarity.

\subsection{Postmerger effects}
\subsubsection{Ejecta properties}
We analyzed the ejecta properties by computing the unbounded ejecta for both the profiles and scalar field masses. To probe the ejecta channel, Fig.~\ref{fig:2dhydroa} shows 2D snapshots of the ejecta showing both the equatorial plane (top panel) and the polar plane (bottom panel) for UP with $\mu = 0.17$ and $M_{\rm tot} = 2.7$, and NF. The equatorial plane, which is typically dominated by shock-driven ejecta, exhibits a noticeable suppression of this material in the presence of the scalar field. This effect is further corroborated by the polar plane plots, which indicate that the scalar field modifies the overall post-merger outflow by suppressing it. Further to support this, we plot the time evolution of volume-integrated unbounded ejecta mass at the AMR level $l=3$ in Fig.~\ref{fig:0dhydroa}. In the top panel, we show the UP for $\mu=0.17; M_{\rm tot} = 2.7$ and NF, along with annotations for tidal-tail (TT) and shock-driven (SD) ejecta mechanisms. In the bottom panel, we show the same for the UP and for scalar field masses $\mu=0.085$ and $\mu=0.17$.  We note that the scalar field enhances the tidal-tail induced ejecta at early times, in turn suggesting that the field reduces the \emph{stellar compactness} due to the gradient pressure. Hence, there is an effective additional pressure contribution that facilitates tidal stripping near contact. In contrast, the shock-driven ejecta is \emph{suppressed}, consequently reducing the overall ejecta. For the lower scalar field mass, the same qualitative pattern persists, but the suppression of the SD ejecta weakens as \(\mu\) decreases, bringing the evolution closer to the vacuum case. It is important to note that these effects are possibly degenerate with other matter contributions, making it unlikely that they are unique signatures of wave DM in the post-merger phase, as well as being dependent on the baryonic EOS. However, our studies show that generally, electromagnetic signatures connected to the outflow of matter from BNS systems could be affected by the presence of the scalar field. 
\newline
The quantities shown in Fig.~\ref{fig:0dhydroa} are extracted at a fixed AMR level within a finite volume, where the observed decrease in ejecta mass over time indicates matter leaving this region. To obtain a more accurate measure of the total ejecta, we instead compute the unbound material crossing a spherical surface of extraction radius $\rm R = 300$ at the finest AMR level; cf.\ Fig~\ref{fig:ejecta}. In the top panel of Fig.~\ref{fig:ejecta}, we plot the ejecta for two different scalar field masses and UP for $\mu = 0.17$ and $\mu = 0.085$, along with the NF simulation. In the bottom panel, we show the unbounded ejecta for both the scalar field profiles OP and UP for $\mu=0.17$ and NF. As discussed before, we find that the presence of the scalar field suppresses the amount of unbound ejecta with respect to the vacuum case, as previously discussed. As the scalar field mass is reduced (i.e., the Compton wavelength increases), the field distribution becomes smoother and the associated gradient–pressure contribution, \( \propto (\nabla\phi)^2 \), lowers the overdensity that builds up around the binary. Due to the reduced accumulation of the
field, the ejecta mass \(M_{\rm ejecta}\) increases, trending back toward the vacuum result. In the bottom panel, we note that both the profiles result in quantitatively similar ejecta mass, both yielding a suppression of the unbound mass by \(\mathcal{O}(60\%)\) relative to the
vacuum simulation, due to a deeper gravitational potential well in the presence of the DM environment on the length scales relevant to the extraction radius. Over time, we note that the ejecta mass reaches a constant plateau. 

\subsubsection{Central rest mass density}
\begin{figure}
    \centering
    \includegraphics[width=\linewidth]{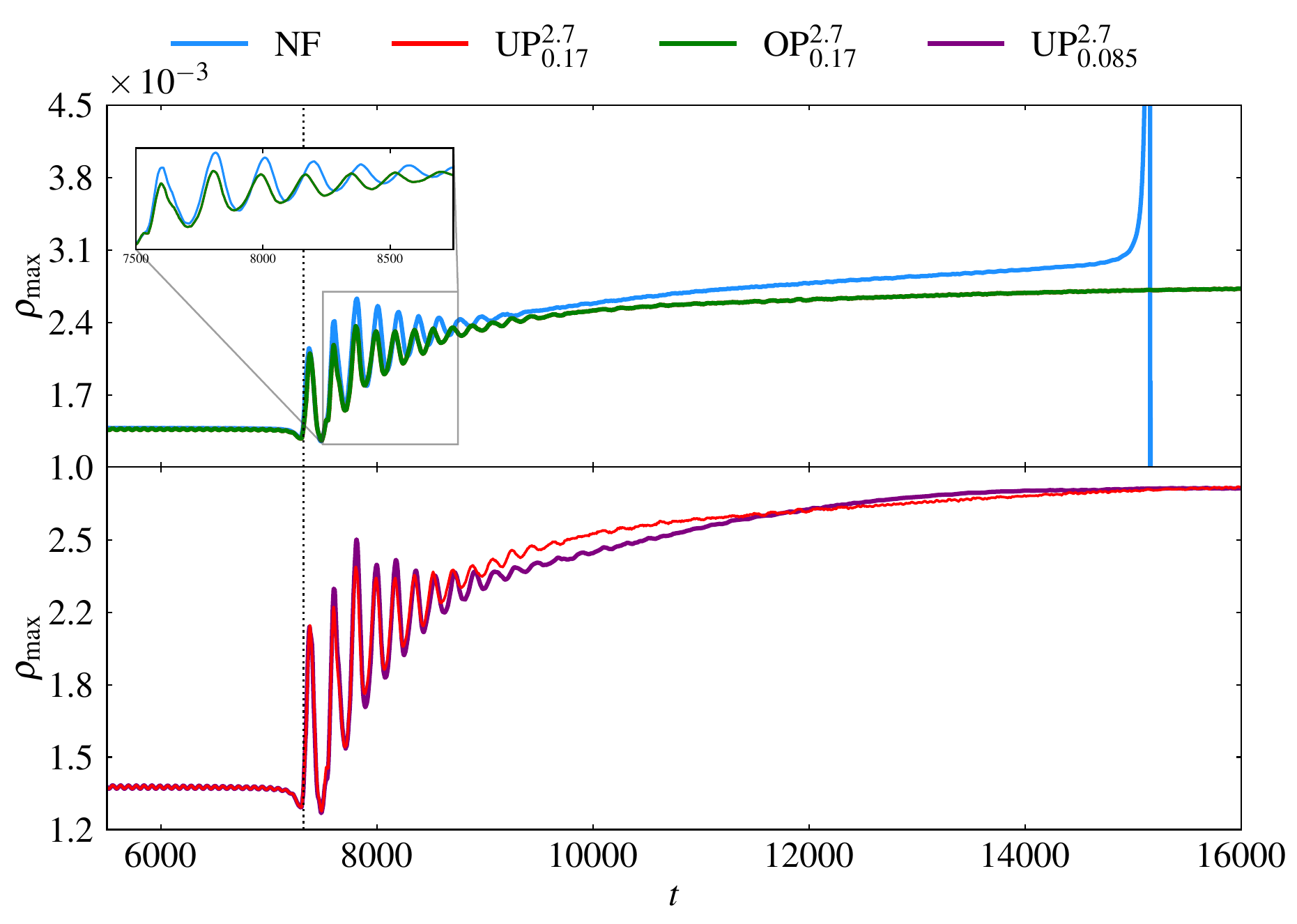}
    \caption{%
Top panel: Evolution of the baryonic central rest–mass density, $\rho_{max}$, through merger and into the post–merger phase for both the scalar field profiles (UP and OP) and for the vacuum reference (NF). The inset zooms the first $t \sim\!10^3$ after merger, showing suppressed oscillations in the presence of the field. Further, the black–hole formation is also \emph{delayed} relative to the vacuum case, as indicated by the energy density growth at $t=15000$.
Bottom panel: The evolution of the maximum baryonic rest--mass density across merger (vertical dotted line) and into the post--merger phase for two scalar--field masses, $\mu=0.17$ (red) and $0.085$ (purple). Lowering the mass of the field increases the amplitude of oscillations in the initial stage of oscillations, followed by a marginally reduced value compared to the $0.17$ case.
In the presence of a scalar field, the post–merger oscillations of $\rho_{max}$ are damped, due to an additional (gradient) pressure that opposes compression of the remnant. }

    \label{fig:rho_alpha_profiles}
\end{figure}

To investigate the effect of scalar field post-merger on the hypermassive remnant, we analyse the baryonic central rest-mass density. During the post–merger, the relevant length scale is set by the remnant size,
\(L_{\rm pm}\sim R_{\rm rem}\), which is smaller than the initial orbital
separation. If \(\mu L_{\rm pm}\lesssim 1\)
(\(\lambda_C=2\pi/\mu\gtrsim L_{\rm pm}\)), the scalar field is not
Yukawa–suppressed on remnant scales and its \emph{gradient} pressure
becomes comparatively more important as seen in the stress–energy tensor in Eq.~\eqref{eqemfield}, the effective “pressure” from gradients scales like \(p_\varphi \sim(\nabla\phi)^2\sim \phi^2/L_{\rm pm}^2\).
In Fig~\ref{fig:rho_alpha_profiles}, we illustrate the evolution of the central density for both the scalar field masses and for both the field profiles along with the vacuum case. Both choices of $\mu$ yield a qualitatively identical evolution: a sharp transition at merger followed by quasi–radial oscillations in $\rho_{\max}$. Quantitatively, the central density is suppressed in the presence of the field; as mentioned, the additional support provided by the gradient pressure resists compression of the remnant. Further, the consequence of the additional gradient pressure is \emph{delayed collapse} of the remnant, while the vacuum simulation forms a BH post-merger. For simulations with the scalar field, we extended the evolution by $t=5000$ beyond the vacuum case, and the remnant did not collapse. This is in contrast to studies involving the particle-like DM admixed with NSs modeled as an ideal fluid, wherein the post-merger remnant collapses sooner than the vacuum case \cite{Emma:2022xjs, Giangrandi:2025rko}. In the lower panel, we show the central density for different scalar field masses ($\rm UP^{2.7}_{0.17}$ and $\rm UP^{2.7}_{0.085}$). The heavier field ($\mu=0.17$) exhibits a slightly smaller oscillation amplitude initially, and a marginally slower growth of $\rho_{\max}$, while the lighter field shows a somewhat deeper first trough; these differences are at the few-percent level, and the curves remain nearly in phase (see inset). A smaller $\mu$ (larger Compton length) confines less of the scalar energy density in the immediate vicinity of the hypermassive neutron star, providing a slightly stronger local pressure support, which in turn resists further compression. In the initial stages of the hypermassive remnant oscillations, the amplitude of the central density is higher in the lower scalar field mass case, closer to what was observed in the vacuum (reference) simulation. Gradually, the counteracting pressure term dominates as the final stages of oscillations are suppressed compared to that of $\mu = 0.17$.

\section{Discussion}\label{sec:discussion}

In this work, we investigated the effects of massive scalar field (wave DM) environments around BNS mergers within full GRHD simulations.
We explored the effect of changes in the scalar–field mass, the total binary mass, and the initial scalar field profiles of the field on the binary and environment co-evolution. 
The DM energy densities are higher than those expected in physically motivated scenarios, with ($\rho_{\varphi} = 1\times 10^{-8}$), which enables us to investigate and quantify the scalar field’s impact on (i) the GW strain and phase, (ii) the amount of unbound ejecta, and (iii) the fate of the post-merger remnant (prompt/delayed collapse), over the relatively short physical timescale of a numerical relativity simulation. The results are summarised below:
\paragraph{GW dephasing:}
We found $\sim 0.3$ radians dephasing during the merger between the vacuum (reference) simulation and the scalar field simulations, with both the profiles showing the quantitatively the same trend for the scalar field mass of $\mu = 0.17$. The lower scalar mass showed lesser dephasing compared to the former, and the higher binary mass simulation resulted in negligible dephasing, as a result of prompt collapse and higher compactness.

\paragraph{Ejecta: Tidal–tail versus shock–driven ejecta:}
There was an $\sim \mathcal O(60\%)$ reduction in the unbounded ejecta mass with the scalar field environment compared to the vacuum case. Once again, the choice of initial profile does not alter the outcome, while variations in the scalar field mass did. In particular, reducing the scalar field mass increases the effective pressure support, driving the system toward a more \emph{vacuum-like} configuration and thereby weakening the gravitational potential well. The scalar field predominantly suppresses the shock-driven component of the ejecta, while only marginally enhancing the tidal-tail contribution. It is also important to note that these effects are degenerate with other matter effects, as well as baryonic EOS-dependent. 

\paragraph{Central density and delayed collapse:}
The presence of a scalar field reduces the central baryonic rest-mass density of the remnant, rendering it \emph{less compact}. 
This results in an earlier post-merger collapse into a black hole in the vacuum case. At the large length scales relevant for computing the ejecta above, the scalar field energy density dominates, thereby deepening the gravitational potential well. In contrast, on scales comparable to the remnant size, the gradient pressure of the field can overcome the energy density. This additional pressure support delays the collapse of the remnant.

\paragraph{Scalar evolution:}
Irrespective of the initial profile chosen, the system appears to tend towards a quasi-equilibrium configuration where the scalar field and the binary co-rotate, with the energy density enhancement close to the NSs about $10^{4}$ times that of the asymptotic value. This is qualitatively similar to the study involving BBHs \cite{Bamber:2022pbs}.  
However, we observed long-lived oscillations in the field profile that depended heavily on the initial configuration, most likely coming from the excitation of additional non-quasi-stationary/higher-order modes, similar to those seen in BBH studies but more pronounced. Although these did not affect the results above, it would be useful to remove them in order to compare the scalar profile with semi-analytical computations \cite{Tomaselli:2024ojz}. 
This would involve solving the coupled Einstein–Klein–Gordon system together with the neutron-star matter equations as was done for binary boson stars in \cite{Siemonsen:2023age}, and for Scalar Gauss-Bonnet gravity in \cite{Nee:2024bur}. This may also help minimize the momentum constraint violation, as we noticed that even with the multigrid solver, we had to fine-tune the iteration parameters to reduce the constraint violation induced by the overlaid field profile. 

In this work, we have shown that wave DM is not dispersed by the BNS and tends to cluster around and within them, in the mass range where there is a correspondence in length scales between the separation and the Compton wavelength, similar to recent results for BBH mergers \cite{Bamber:2022pbs, Aurrekoetxea:2023jwk}. However, even at very high densities, scalar matter does not significantly affect the dynamics or merger signatures of NSs. While the simulations included only around ten orbits of the binary close to merger, the realistic inspiral timescale is of the order of Gyr. During this long early-inspiral phase, a higher fraction of DM could, in principle, accumulate around the NSs, potentially enhancing its effects within the sensitivity band of GW detectors \cite{Abac:2025saz, KAGRA:2023pio}. Nevertheless, even in such scenarios, the signatures identified here would remain difficult to disentangle from other effects, resulting in strong degeneracies. Therefore, detecting any accumulation of wave DM using NSs will remain challenging even with upcoming detections and future observational facilities. 

One case in which the effects might be enhanced is that of neutron star–black hole mergers, where the scalar field could have undergone amplification through superradiance around the black hole during its history, resulting in higher densities. Further, since the main candidates for wave DM are axions or ALPs, one should account for the effects of their expected self-interactions, or couplings to matter and gauge fields. These effects, as well as some proposed production mechanisms, can lead to larger accumulation and novel dynamics in and around compact objects \cite{Marsh:2015xka, Hook:2017psm, Kumamoto:2024wjd, Iwamoto:1984ir}. Complementary studies exploring different mass ranges of ALPs have placed constraints based on their nucleonic coupling and emission from supernovae cores \cite{Lella:2023bfb}, supernovae remnants \cite{Beznogov:2018fda}, cooling of NSs \cite{Fiorillo:2025zzx}, and pulsars \cite{Noordhuis:2023wid}, see \cite{Caputo:2024oqc} and references therein.
Constraints on DM self-interactions arising from the Bullet Cluster \cite{Robertson:2016xjh} depend on the particle mass, and significant interactions could still play a role on smaller scales \cite{Berezhiani:2015bqa, Berezhiani:2019pzd}.
In such scenarios involving additional couplings, the dynamics of scalar interactions and the formation of bound states may give further observational signatures.

\section{Acknowledgements}
We acknowledge useful conversations with Josu Aurrekoetxea, Jamie Bamber, Francesca Chadha-Day, and Giovanni Maria Tomaselli.
KC acknowledges support from the Simons Foundation International and the Simons Foundation through Simons Foundation grant SFI-MPS-BH-00012593-03, a UKRI Ernest Rutherford Fellowship (grant number ST/V003240/1), and an STFC Research Grant ST/X000931/1 (Astronomy at Queen Mary 2023-2026). We acknowledge funding from the European Union (ERC, SMArt, 101076369). Views and opinions expressed are those of the authors only and do not necessarily reflect those of the European Union or the European Research Council. Neither the European Union nor the granting authority can be held responsible for them. Computations for this paper have been performed on the DFG-funded research cluster Jarvis at the University of Potsdam (INST 336/173-1; project number: 502227537). 
\bibliography{refs.bib}

\appendix
\section{Convergence tests and Diagnostics}\label{sec:Appendix}
\subsection{Angular momentum exchange}\label{subsec:ADM}
Following \cite{Clough:2021qlv}, we outline the equations for the conservation law for angular momentum along the vector $\zeta^\nu$. 
We need to check for the conservation equation,
\begin{equation}
    \partial_t \int_\Sigma Q \ dV = \int_{d\Sigma} F\cdot\hat{n} \ dA + \int_{\Sigma} S \ dV,
\end{equation}
including the source term $S$, which describes the exchange of the charge between matter and curvature, and an additional term in the conservation equation, as $\zeta^\nu$ is not a Killing vector. The source term quantifies the way the matter (scalar field in this case) extracts momentum from the binary, and corresponds to gravitational forces in the
Newtonian limit \cite{Aurrekoetxea:2023jwk, Clough:2021qlv}. For angular momentum, $\zeta^\nu = \partial_\phi = (1, -y, x, 0)$, we get,
\begin{align}
Q_{\{\phi\}} &= S_i \xi^i_{\{\phi\}} \\
F_{\{\phi\}} &= - N_i \beta^j S_j \xi^j_{\{\phi\}} + \alpha N_i S_j \xi^j_{\{\phi,r\}} \\
S_{\{\phi\}} &= \alpha S^\mu_{\nu} \partial_0 \xi^\nu_{\{\phi\}} + \alpha S^\mu_{\nu} {}^{(3)} \Gamma^\nu_{\mu \sigma} \xi^\sigma_{\{\phi\}} \\ \nonumber
&\quad - S_\nu \beta^i \partial_i \xi^\nu_{\{\phi\}} + S_\nu \xi^\mu_{\{\phi\}} \partial_\mu \beta^\nu - \rho \xi^\mu_{\{\phi\}} \partial_\mu \alpha
\end{align}
\noindent where the time derivative of the 3-metric is given by:
$\partial_t \gamma_{ij} = - 2 \alpha K_{ij} + D_i \beta_j + D_j \beta_i$, 
and the quantities \(\partial_\mu \xi^\nu\) are zero except for:
$\partial_x \xi^y = 1, \quad \text{and} \quad \partial_y \xi^x = -1$.
\begin{figure}
    \centering
    \includegraphics[width=\linewidth]{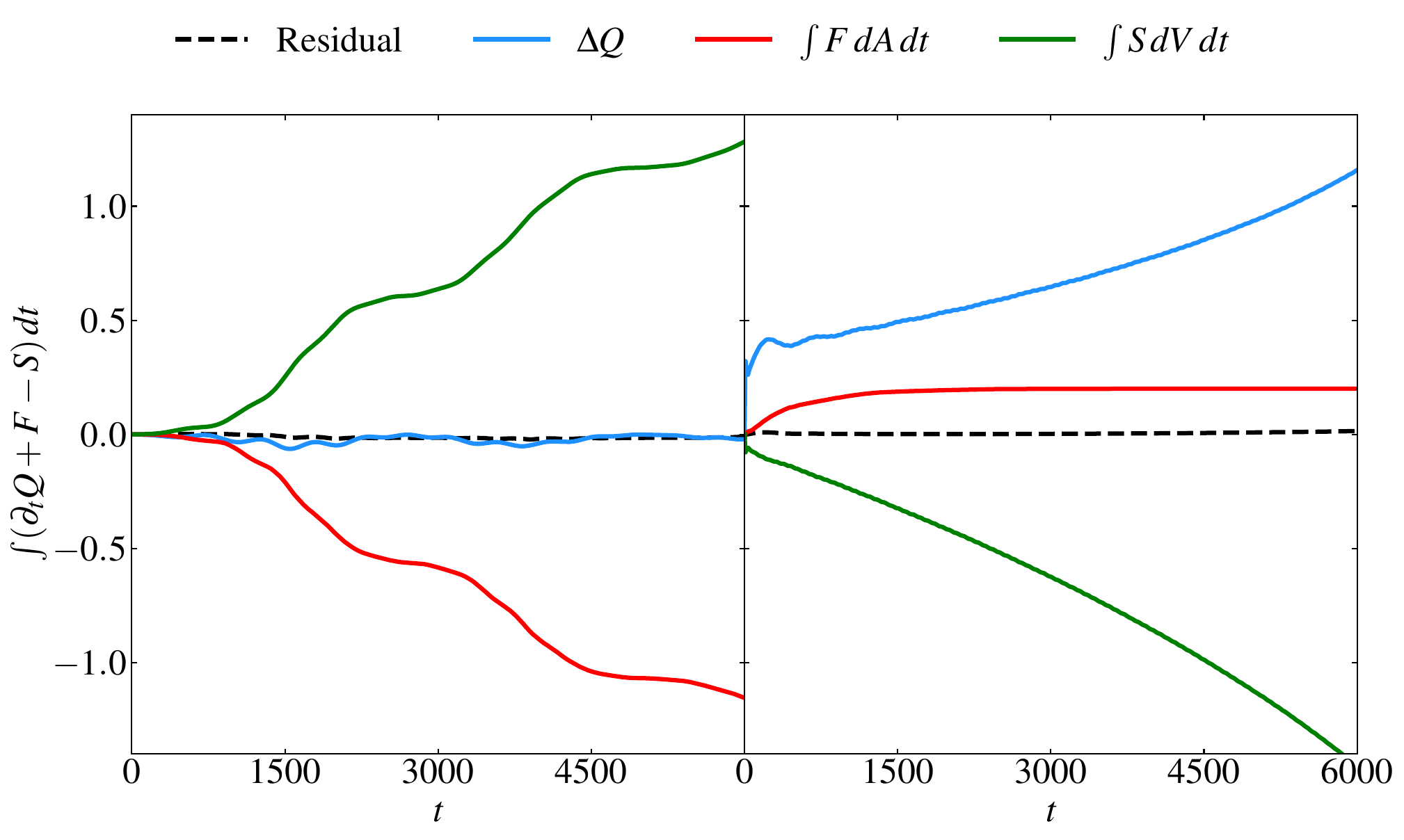}
    \caption{Angular momentum exchange of the scalar field (Left) and for the NSs (Right), integrated with time extracted at a radius $\rm R = 40$. The green line shows the change in each respective source term for the field and the stars; the red line represents the total flux of the charge from an outer boundary of the extraction sphere. The blue line represents the evolution of the charges, and the black dashed line is the residual. On the left, we see that for the field, the flux is balanced by the variation of the corresponding source term, suggesting that the extraction of angular momentum occurs via the background curvature. Right: For the NSs, the net flux contribution from the matter sector saturates, while both the charge and the source term exhibit an approximately linear growth at later times, complementing each other.}
    \label{fig:adm}
\end{figure}
\begin{figure}[!htbp]
    \centering
    \includegraphics[width=\linewidth]{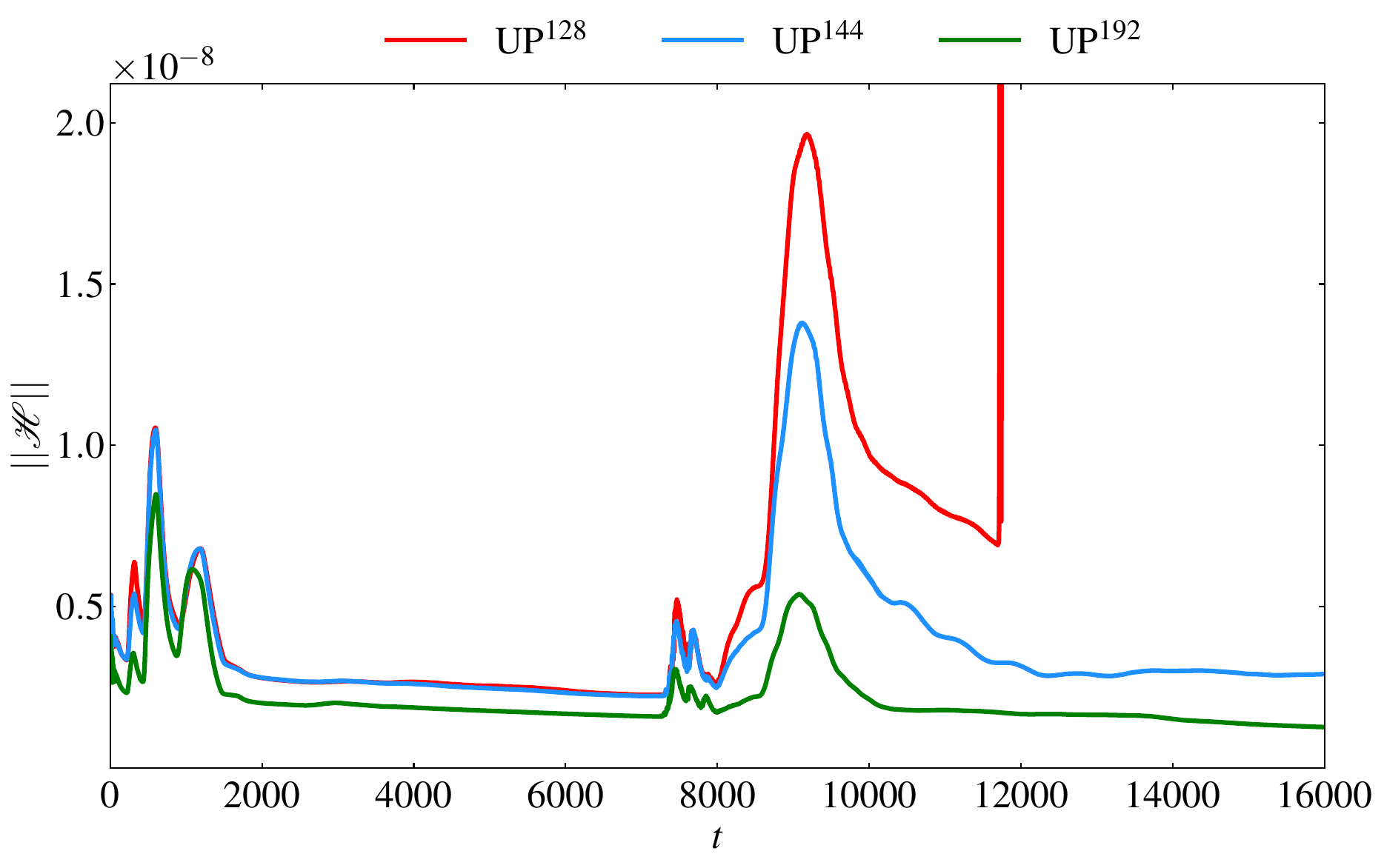}
    \includegraphics[width=\linewidth]{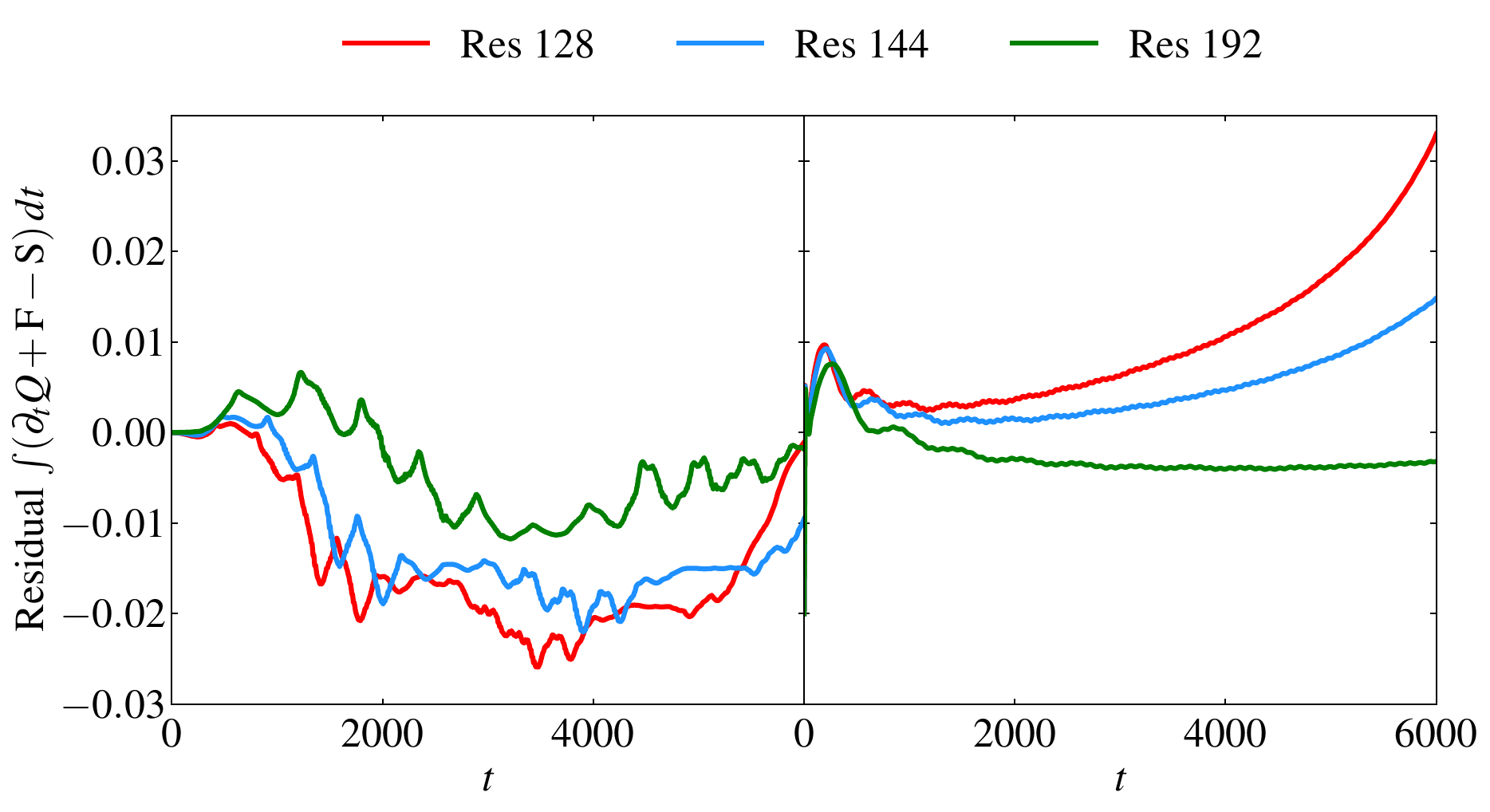}
    \includegraphics[width=\linewidth]{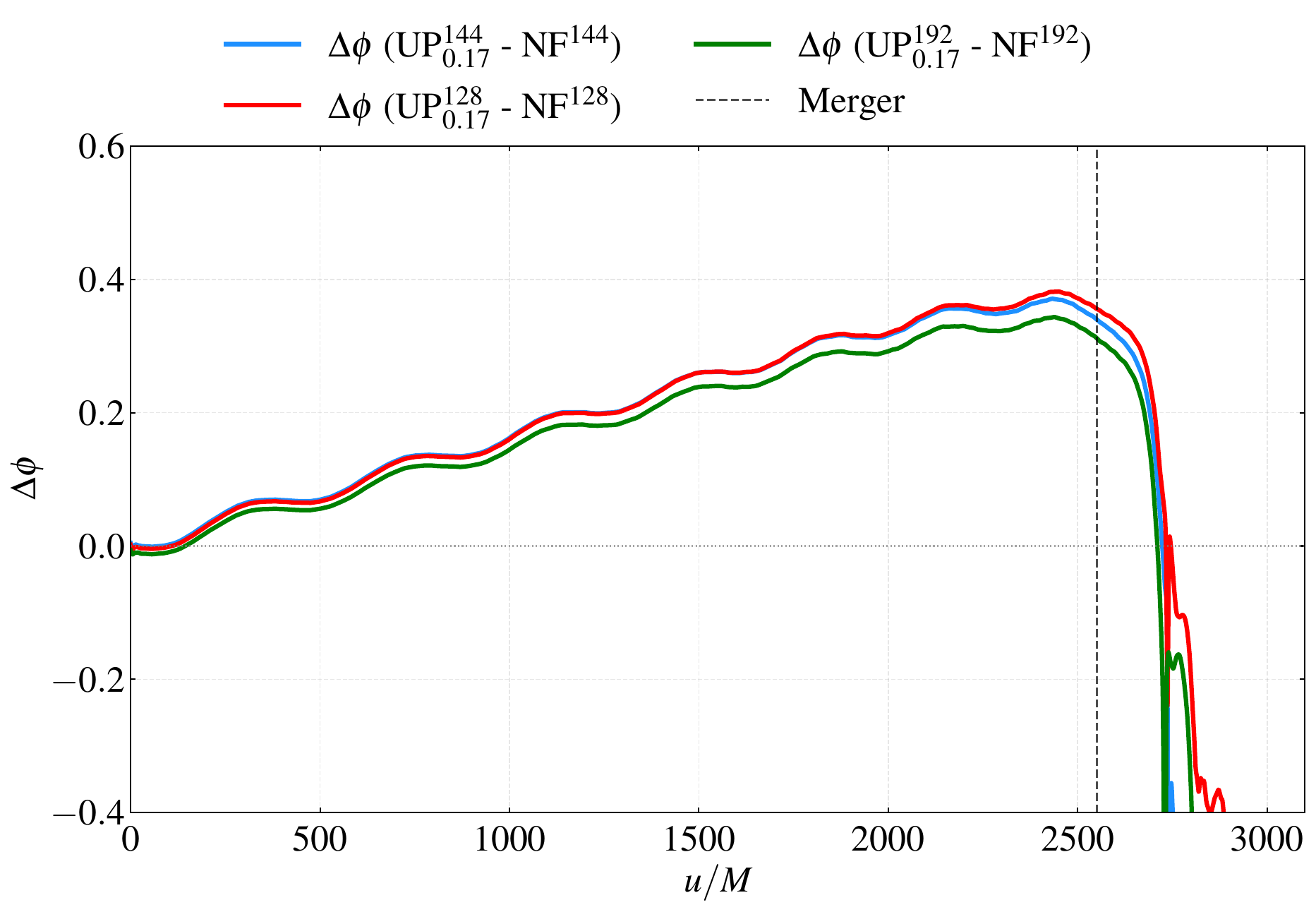}
    \caption{Top: Time evolution of the L2-norm of the Hamiltonian
constraint for the $M_{\rm tot}= 2.7$ simulation, and Uniform profile with $\mu = 0.17$ for all three resolutions, with $\rm UP^{128}$ being the coarsest and $\rm UP^{192}$ being the finest. Middle: Convergence of residual of the angular exchange for the scalar field (left) for UP and $\mu = 0.17$ and NSs (right). We note that the residual tends to zero as we increase the resolution. Bottom: Convergence of the GW phase difference between the UP and NF for different resolutions and for $\mu = 0.17$.}
    \label{fig:allham}
\end{figure}
In Fig.~\ref{fig:adm}, we present the time-integrated quantities associated with the angular momentum conservation for both the scalar field (left panel, UP) and the NSs (right panel). For the scalar field, the evolution of the corresponding charge (blue), the total flux across a finite extraction sphere (red), and the source terms (shown in green) are displayed. Since the scalar field initially carries no net angular momentum (i.e., the charge), its subsequent angular momentum is entirely due to scalar radiation generated through interaction with the binary. The flux is balanced by the variation of the corresponding source term, suggesting that the extraction of angular momentum occurs via the background curvature, effectively, the dynamical friction.

In the case of the NSs (right panel), the net flux contribution from the matter sector quickly saturates (appearing constant once integrated in time), while both the charge and the source term exhibit an approximately linear growth at later times. This behavior can be attributed to resolution effects: at higher resolutions (see the middle panel in Fig.~\ref{fig:allham}), the magnitudes of these terms reduce, indicating that they primarily arise from numerical artifacts associated with the outer stellar layers interacting with the binary’s angular momentum and gravitational interaction.
\subsection{Convergence tests}\label{subsecConv}
In Fig.~\ref{fig:allham}, we present convergence results for UP with a scalar field mass of $\mu = 0.17$ and $M_{\rm tot} = 2.7$. The top panel shows the convergence of the Hamiltonian constraint, the middle panel displays the convergence of the angular momentum exchange terms, and the bottom panel illustrates the GW phase difference for three different resolutions: 128, 144, and 192. For the Hamiltonian constraint, the violation decreases with increasing resolution with approximately second order. The lowest-resolution run collapses to a black hole around $t \approx 12000$, whereas the medium and high-resolution runs remain stable. In the middle panel, the angular momentum residuals for both the NSs and the scalar field approach zero as the resolution increases. Finally, in the bottom panel, the GW phase differences converge toward the same value as the resolution is increased.
\subsection{Noether charge}\label{subsecnoether}
The scalar field Lagrangian is invariant under a global $U(1)$ symmetry transformation $\varphi \rightarrow \varphi e^{i\epsilon}$. This symmetry, following Noether's theorem, gives rise to a conserved current $J_{\phi}$, and a corresponding charge $Q_{\phi}$.
The current density can be expressed in terms of the action of the field as 
\begin{equation}
    J^{\mu}_{\phi} = \frac{\delta S_{\varphi}}{\delta (\partial_\mu \varphi)} \delta\varphi, \label{eqJ}
\end{equation}
which is conserved hence $\nabla_i J^i_{\phi} = 0$. The conserved Noether charge is the time component of the current density, hence $Q_{\phi} = \int_{\Sigma_t} J^t$. 
Considering the scalar field action Eq. (\ref{eqaction}) and using Eq.~\eqref{eqJ}, we get,
\begin{equation}
    J^{\mu}_{\phi} = i g^{\mu\nu} \left(\varphi \partial_\nu \bar{\varphi} - \bar{\varphi}\partial_\nu \varphi \right), \label{noethercurrent}
\end{equation}
and to compute the charge, we expand the above and substitute in the metric components, 
\begin{align}
    J^{t}_{\phi} = i\left[g^{tt}(\varphi \partial_t \bar{\varphi} - \bar{\varphi}\partial_t \varphi) + g^{ti} (\varphi \partial_i \bar{\varphi} - \bar{\varphi}\partial_i \varphi)\right],
\end{align}
and we arrive at,
\begin{equation}
    Q_{\phi} = \int_{\Sigma_t} \left(i[\bar{\varphi} \Pi - \bar{\Pi}\varphi] \right) \ d^3x. \label{noethercharge}
\end{equation}
\section{Multigrid formalism}\label{subsecMG}
\begin{figure}
    \centering
    \includegraphics[width=\linewidth]{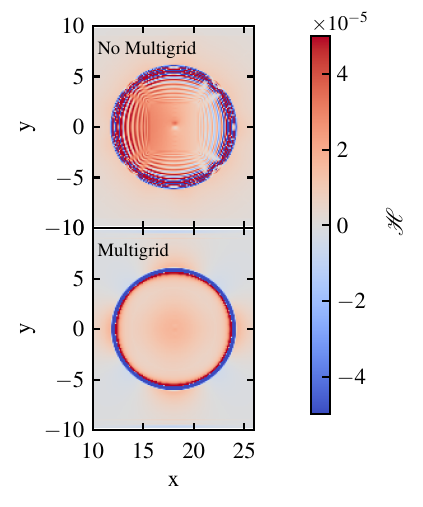}
    \caption{2D snapshot of the Hamiltonian constraint at $t=0$ around a single star at the finest refinement level. The top panel shows the result without the multigrid solver, while the bottom panel shows the result with multigrid, showing comparatively smoother and reduced constraint violations around and within the star.}
    \label{fig:2dham}
\end{figure}
In the 3+1 framework, the  Einstein constraint equations are given by
\begin{align}
    R + K^2 - K_{ij} K^{ij} = 16\pi \rho_t, \label{hamcons}\\
    D_{j} K^{ij} - D^i K =  8\pi S^i_t, \label{momcons}
\end{align}
where the former is the Hamiltonian constraint and the latter are the Momentum constraint equations. $D_i$ is the three-dimensional covariant derivative after orthogonal projection to the normal vector of the hypersurface. The subscript 't' in $\rho_t$ and $S^i_t$ indicates energy density and momentum density, respectively, including both, the scalar and baryonic matter contributions. Decomposing the two, we get $\rho_t = \rho_m + \rho_\varphi$; $S^i_t = S^i_{m} + S^i_{\varphi}$. 
\paragraph{NSs}
For a perfect fluid with rest-mass density $\rho$, pressure $p$, specific enthalpy $h$, 3-velocity $v^i$, and Lorentz factor $W = (1-v^2)^{-1/2}$, we have, 
\begin{align}
\rho^m &= \rho h W^2 - p, \\
S^i_{m} &= \rho h W^2 v^i .
\end{align}
\paragraph{Scalar field}
As mentioned in \ref{subsec:KG}, the time derivative of the scalar field is
\begin{equation}
\Pi \equiv \frac{\sqrt{-g}}{\alpha^2}
\left( \partial_t \varphi - \beta^i \partial_i \varphi \right),
\end{equation}
and the scalar-field contributions to the ADM sources are,
\begin{align}
\rho^{\varphi} &=
\frac{1}{2} \left(|\Pi|^2
+ \gamma^{ij} \partial_i \varphi \, \partial_j \varphi^*
+ \mu^2 \varphi \varphi^*
\right), \\
S^i_{\varphi} &=
\frac{1}{2} \gamma^{ij}
\left(
\Pi^*  \partial_j \varphi
+ \Pi  \partial_j \varphi^*
\right).
\end{align}

In the Extended conformal thin-sandwich formulation (XCTS) \cite{Pfeiffer:2002iy}, we decompose the spatial metric $\gamma_{ij}$ into a spatial conformal metric with a conformal factor $\psi$ taking the form $\bar{\gamma}_{ij} = \psi \gamma_{ij}$. Along with the spatial metric, we also decompose the extrinsic curvature as 
\begin{align}
    K_{ij} \equiv -\frac{1}{2} \mathcal{L}_{n} \gamma_{ij} = A_{ij} + \frac{1}{3} \gamma_{ij} K, 
\end{align}
where $K = K^{i}_i$ is the trace of the extrinsic curvature, and the rescaled traceless part $A_{ij}$ is
\begin{align}
    A_{ij} = -\frac{\psi^4}{2\alpha} \left[\partial_t \bar{\gamma}_{ij} - (\bar{L} \beta)_{ij} \right],
\end{align}
here,
\begin{align}
    (\bar{L}\beta)_{ij} = \bar{D}_i \beta_j + \bar{D}_j \beta_i - \frac{2}{3} \bar{\gamma}_{ij} \bar{D}_m \beta^m,
\end{align}
is the traceless part of $\mathcal{L}_{\beta} \bar{\gamma}_{ij}$, and $\bar{D}$ is the covariant derivative with respect to the conformal spatial metric $\bar{\gamma}_{ij}$. 
We can simplify the system of equations assuming spatial conformal flatness (CFA) and with maximal slicing such that,
\begin{equation}
    K = \partial_t K = 0; \ \bar{\gamma}_{ij} = f_{ij}, \ \partial_t\bar{\gamma}_{ij} = 0,
\end{equation}
where $f_{ij}$ is the flat space metric, and we get,
\begin{align}
    \partial^i \partial_i \psi = -\frac{1}{8} \psi^5 \left(A_{ij} A^{ij} + 16\pi \rho_t\right) \label{eqpsi}, \\ 
    \partial^j \partial_j \beta^i + \frac{1}{3} \partial^i \partial_j \beta^j = 2\psi^{10} A^{ij} \partial_j(\alpha \psi^{-6}) + 16\pi \alpha \psi^4 S^i_t \label{eqbeta}, \\
   \partial^i\partial_i(\alpha\psi) = \alpha \psi^5 \left(\frac{7}{8} A_{ij} A^{ij} + 2\pi(\rho_t+2S_t)\right), \label{eqapsi}   
\end{align}
Fig.~\ref{fig:2dham} shows the 2D snapshot of the Hamiltonian constraint at $t=0$ around a single star in the binary system with and without resolving the constraints using the multigrid solver. We note that with multigrid, although there are constraint violations, particularly around the surface of the star, they are reduced and \textit{smoother} compared to the simulations without multigrid, as shown in the top panel. 
\end{document}